\documentclass{vldb}
\usepackage{algorithm}
\usepackage{algpseudocode}
\usepackage{graphicx}
\usepackage{balance} 
\usepackage{subcaption}
\usepackage{url}
\usepackage{balance}

\begin{document}
%

\title{Processing Analytical Workloads Incrementally}

%
%
%
%
%

\numberofauthors{3} 
%
\author{
%
%
\alignauthor 
Priyank Gupta\\
       \affaddr{University of Toronto}\\
       \email{priyank@cs.toronto.edu}
\alignauthor 
Nick Koudas\\
       \affaddr{University of Toronto}\\
       \email{koudas@cs.toronto.edu}
\alignauthor 
Europa Shang\\
       \affaddr{University of Toronto}\\
       \email{europa@cs.toronto.edu}
\and
\alignauthor 
Ryan Johnson\\
       \affaddr{University of Toronto}\\
       \email{ryan.johnson@cs.utoronto.ca}
\alignauthor Calisto Zuzarte\\
       \affaddr{IBM Toronto}\\
       \email{calisto@ca.ibm.com}
}

\maketitle
\begin{abstract}
Analysis of large data collections using popular machine learning and statistical algorithms has been
a topic of increasing research interest. A typical analysis workload consists of applying an algorithm to build a model
on a data collection and subsequently refining it based on the results.

In this paper we introduce model materialization and incremental model reuse as first class citizens in
the execution of analysis workloads. We materialize built models instead of discarding them in a way that can be reused in subsequent computations. At the same time we consider manipulating an existing model (adding or deleting data from it) in order to build a new one. We discuss our approach in the context of popular machine
learning models. We specify the details of how to incrementally maintain models as well as outline the
suitable optimizations required to optimally use models and their incremental adjustments 
to build new ones. We detail our techniques for linear regression, naive bayes and logistic regression and
present the suitable algorithms and optimizations to handle these models in our framework.

We present the results of a detailed performance evaluation, using real and synthetic data sets. Our experiments
analyze the various trade offs inherent in our approach and demonstrate vast performance benefits.
\end{abstract}

\section{Introduction}

Analytics on large collections of data is a topic of vast interest in recent years. Although analysis of data was always central in the data management community, the prevalence of various machine learning and statistical systems/packages has corroborated to the interest. As a result several recent lines of research across communities aim to engineer popular machine learning techniques both at the algorithmic as well as the systems level to scale in large data collections \cite{Apache:MLLib,Pivotal:MADLib,Zinkevich:PSGD,Mann:SGD}.

Data analytics tasks however, are rarely run in isolation. Typically an analysis workload consists of applying an algorithm (e.g., machine learning algorithm or statistical operation) on a large data set building a model and subsequently refine the operation based on the results of previous steps. For example consider building a model (e.g., regression operation) on a data set produced for the first two weeks of a month (e.g., sales data as it relates to various traffic parameters and promotions activities on a web site).
Based on the results of the operation (e.g., regression parameters, error, etc) one decides to run an additional regression operation for the data set representing the entire month. Alternatively during a data exploration task, one creates a data model for a year worth of data collected for a service, only to decide to drill down and build a model for the second month of the year that seems to present an anomaly for the given model fit.

It is evident that analysis tasks can be part of an analysis workload and rarely run in isolation. Moreover, exploratory tasks, may involve extending or refining previously completed tasks. As a result, this behavior reveals certain dependencies among the steps of an analysis workload. Such dependencies expose opportunities for work sharing across tasks. 
For example one may be able to reuse the model for the first two weeks
of the month instead of building the model for the entire month from scratch. Such reuse could be 
achieved by incrementally updating the current model with additional data. Alternatively if the model for the subsequent two weeks of the month is available, the desired model for the month could be build by combining the two models as opposed building it from scratch. Such an option is advantageous as the models are already build and one simply derives a new one without the need to access possibly large collections of data. In a similar fashion we may be able to reuse the model build for a month to derive the model for the first two weeks of the month by removing the last two weeks worth of data from the model, instead of building the desired model from scratch.

These examples reveal two basic observations that we explore further in this paper. First analysis workloads consisting of multiple modelling tasks are amenable to work sharing across tasks. In particular one may be able to reuse models previously build on a data set in order to derive new models on demand. Second, incremental updates (inserting or deleting data) is an operation that may aid to derive a new model from an existing one. It is natural to expect that some models would enable work sharing easier than others. Some models for example may allow us to derive a new model by "extending" (with new data) or "shrinking" (removing data) the current model and still derive the exact same model we would have derived by building it from scratch utilizing base data. Some other models could allow us to do this only approximately. At the same time from a performance standpoint it may not always be beneficial to utilize an existing model and derive a new one by adding or deleting data from it. We expect that in some cases utilizing an existing model to derive a new one may be beneficial (we may be able to build the model much faster) but in some other cases, building the model from scratch is the best (faster) option.

Currently, systems that enjoy vast attention and are utilized for data analysis tasks (e.g., R \cite{R:Analytics}) do not take advantage of such dependencies and inherent relationships across operations of a data analytics workload. An analyst has to be aware of work sharing opportunities as well as optimization opportunities and express them (in code) explicitly which is not an ideal solution. 

In this paper we initiate a study to explore these possibilities. We introduce {\em model materialization} and {\em incremental model reuse} as first class citizens in the execution of an analytical workload. By model materialization we mean that a model can be stored after it is build in order to be considered when generating other models. Since a model requires some space to store it, we incur a storage cost but we aim to offset such costs with increased performance in executing subsequent operations. By incremental model reuse we mean that during the decision to build a model required by an analyst, we consider models previously build as candidates to generate the model. Thus, we decide whether we should reuse existing models and/or adjust them incrementally or build the model from scratch. The decision is typically based on performance and we aim to make the choice that results in building the model fastest. Towards this goal we adopt a cost model that aids in this decision; we develop the suitable optimization frameworks that decide which models to use and the suitable action to take with the objective of producing the resulting model with the smallest cost.

More specifically in this paper we make the following contributions:
\begin{itemize}
\item We introduce {\em model materialization} and {\em incremental model reuse} as frameworks to be considered during the execution of an analysis workload.
\item Using linear regression and Naive Bayes as examples, we demonstrate how these common models can be casted in our framework. More specifically we establish that incremental model reuse and model materialization offer large performance benefits, while guarantying that models are constructed without loss of accuracy.
\item We introduce an algorithm that given a collection of materialized linear regression/naive bayes models, chooses the best models to reuse and also the suitable operations in order to modify them deriving the desired target model with minimal cost.
\item Using logistic regression as an example, we demonstrate that incremental model reuse and model materialization offer large performance benefits while guarantying that models are constructed with quantifiable loss in accuracy.
\item We introduce an algorithm that given a collection of logistic regression models, chooses the best models to reuse and the suitable operations in order to modify them deriving the desired target model with minimal cost.
\item We present the results of an extensive performance comparison demonstrating the performance benefits of our approach under varying parameters of interest.
\end{itemize}

This paper is organized as follows: Section \ref{sec:background} presents introductory material and basic notation. Section \ref{sec:incremental} demonstrates incremental manipulation of linear regression and naive bayes models, followed by Section \ref{sec:logistic} that treats the case of logistic regression models. Section \ref{sec:opt} introduces our optimization framework followed by Section \ref{sec:exp} that details and empirical evaluation of the proposal. Section \ref{sec:related} discusses related work and Section \ref{sec:conc} concludes the paper.

\section{Background}

\label{sec:background}
We provide basic notation and a brief introduction to the techniques we adopt to showcase our overall approach.
A more detailed description of the algorithms is available elsewhere \cite{Bishop:PRML, Murphy:MLPP}

\renewcommand{\vec}[1]{\mathbf{#1}}

\subsection{Linear Regression}
Linear regression is modelling the relationship between a scalar dependent variable and one or more independent variables. Consider a data set of $n$ records; each record $x$ is a $d$-dimensional feature vector of independent variables denoted by $\vec{x_i}$ and a target dependent variable $y_i \in $ $\mathbb{R}$. Generally, a linear regression takes the following form :
\begin{displaymath} y_i =  w^T\vec{x_i} + {\epsilon}_i \end{displaymath} 
where $w$ is the weight vector which is estimated and ${\epsilon}_i$ is an error term. Usually, the weight parameters are learned by minimizing sum of squared errors. 
A $L_2$-regularization term is added to avoid over-fitting of the model.
The solution thus obtained has a closed form and is represented as : \begin{equation} \vec{w} = ({\vec{X}}^T\vec{X} + \lambda I)^{-1}(\vec{X}^T\vec{y}) \end{equation} 
$\vec{X}$ is a $n\times d$ matrix of the input vectors, $\vec{y}$ is a $n\times1$ matrix of the target values and $\lambda$ is the regularization parameter.

\subsection{Naive Bayes Classifier}
\label{sec:naive-bayes}
Naive Bayes classifiers are simple probabilistic models assuming pair-wise independence of features given the class label. 
Albeit simple, Naive Bayes models perform very well in classification problems\cite{Zhang:NB}.
Given a class variable $Y$ and a set of predictor variables $x_1,...,x_d$ Bayes theorem states that 
\begin{displaymath} P(Y=c|x_1,....,x_d) =  \frac{P(Y=c).P(x_1,....,x_d|Y=c)}{P(x_1,....,x_d)} \end{displaymath} 

Under the naive assumption and given that ${P(x_1,....,x_d)}$ is constant for a particular training set we can conclude that 

\begin{displaymath} P(Y=c|x_1,....,x_d) \propto  P(Y=c).\prod_{i=1}^{d}P(x_i|Y=c) \end{displaymath} 

$P(Y=c)$ can be calculated from training data by maximum likelihood estimation. The class probability $P(Y=c)$ is simply the relative frequency of class $c$ in the training set, $P(Y=c)=  N_c/N$ where $N_c$ is 
number of training example which have class $c$ and $N$ is the total number of training examples.

Depending upon the choice of distribution for the conditional density $P(x_1,....,x_d|Y=c)$ we have variations of the Naive Bayes classifier. A popular choice in the case of real valued features is the Gaussian distribution. 
\begin{displaymath}  P(x_1,....,x_d|Y=c) = \prod_{j=1}^{d}\mathcal{N}(x_j|\mu_{jc},\sigma_{jc}^{2}) \end{displaymath} where $\mu_{jc}$ is the mean of feature $j$ in samples with class label as $c$ and $\sigma_{jc}^{2}$
is its variance. This is often referred to as \textit{Gaussian  Naive  Bayes}.
In case of categorical features the multinomial distribution is a preferred choice for conditional density. The distribution is parametrized by vectors 
$ \theta_c = (\theta_{c1},...,\theta_{cd})$ for each class, $d$ is the dimension of the feature vector and $\theta_{ci}$ is the probability $P(x_i|c)$ of feature $i$ appearing in sample belonging to class $c$. 
\begin{displaymath}  P(x_1,....,x_d|Y=c) = (\sum_i^dx_i)!\prod_{i=1}^{d}\frac{\theta_{ci}^{x_i}}{x_i!} \end{displaymath}
$\theta_{ci}$ can be calculated by a smoothed version of maximum likelihood estimation. 
\begin{displaymath} \theta_{ci} = \frac{N_{ci}+1}{N_c + d} \end{displaymath}
where $N_{ci} = \sum_{j=1}^n x_i^{(j)}[Y=c]$
, $N_c = \sum_{i=1}^d\sum_{j=1}^n x_i^{(j)}[Y=c]$ and $n$ is the total number of points in the training set. These counters are computed for each class in the training data.
 
\subsection{Logistic Regression}

Logistic regression is a linear classifier belonging to the family of Generalized Linear Models \cite{Bishop:PRML}. Let $y$ denote a class variable and $x$ represent a feature vector, then  Logistic Regression can be formally represented as an optimization problem minimizing a
loss function to identify the model parameters. The loss function has the following form
\begin{equation} \label{eqn:logistic} 
F(w) = \frac{1}{n}\sum_{i=1}^n L(w;x^{(i)},y^{(i)}) + \lambda R(w)
\end{equation}

A very common choice for function $L$ in logistic regression is the cross entropy loss function :
\begin{displaymath}L(w;x^{(i)},y^{(i)}) = 	y^{(i)} log h_w(x^{(i)}) + (1-y^{(i)}) log (1- h_w(x^{(i)})) \end{displaymath}
and regularization function $R(w) = \| w \| ^ 2 $. Here $h_w(x)$ is the logistic function $h_w(x) = \frac{1}{1+ e^{-w^Tx}}$. 

The Stochastic Gradient Descent(SGD) algorithm \cite{Murphy:MLPP} is used to optimize the loss function to determine the model parameters. SGD initializes the model parameter $w$ to some
$w_0$ and then updates the parameter as 
\begin{displaymath} w \leftarrow w - \alpha \nabla F_i(w) \end{displaymath}

where $\alpha$ is the learning rate and $\nabla F_i(w)$ is the gradient of the convex loss function just using the $i^th$ sample. Stochastic gradient descent requires a single pass on the data 
to converge.

\newtheorem{theorem}{Theorem}
\begin{sloppypar}
\section{An Incremental Approach} \label{sec:reg}
\label{sec:incremental}
We now demonstrate how model materialization and incremental model reuse can be supported in each of the types of models we consider. We discuss how one can combine two models on different data sets to produce a new model on the union of the data sets. We also discuss how an existing model can be manipulated (by adding or removing data) to produce a new one. Formally, let $M_1$ be a model on data set $D_1$ and $M_2$ is the model on data set $D_2$. We assume that the data sets $D_1$ and $D_2$ have the same properties.
We discuss two machine learning models described in the previous section, Linear Regression and Naive Bayes. 

\subsection{Model Materialization}

A typical machine learning model is characterized by its parameters. In order to support incremental updates to a given model extra information has to be maintained depending on the model. We show that while materializing a model we can also materialize extra information that would be sufficient in supporting incremental updates. This information varies across different types of models as discussed further in this section.

\subsubsection{Linear Regression}
\label{subsec:lr}

Let $D$ be a data set of $n$ points and let $M$ represent a machine learning model build on this data set. 

Parameters for a linear regression are provided by Equation 3. The equation can be considered as a combination of two terms $ A = X^TX $ and $ B = X^Ty $. 
Simplifying the terms 
\begin{displaymath} X^TX = \begin{bmatrix}
\sum_{j=1}^n x_1^{(j)} x_1^{(j)}  & \dots & \sum_{j=1}^n x_1^{(j)} x_d^{(j)} \\
\vdots & \ddots & \vdots \\
\sum_{j=1}^n x_d^{(j)} x_1^{(j)}  & \dots & \sum_{j=1}^n x_d^{(j)} x_d^{(j)} \\
 \end{bmatrix} 
\end{displaymath}

\begin{displaymath} X^TY = \begin{bmatrix}
\sum_{j=1}^n x_1^{(j)} y^{(j)}  \\
\vdots  \\
\sum_{j=1}^n x_d^{(j)} y^{(j)}  \\
 \end{bmatrix} 
\end{displaymath}

where $A$ is a $d \times d$ matrix and each term is the sum product of any two features of the feature vector over the $n$ training samples. $X^Ty$ is a $ d\times 1$ matrix where each term is the sum product of the features and the target values. We will maintain matrix $A$ and $B$, along with the model parameters while building a model. Thus we end up maintaining $ d^2+ d $ extra values. It is important to note that the amount of extra information we have to maintain is independent of the number of training samples ($n$). Given that we have both the components $A$ and $B$ we can compute the model parameters at any point using equation \ref{eqn:logistic}. Later on we will show how we can support incremental updates to Linear Regression model utilizing this information.

\subsubsection{Naive Bayes}
\label{subsec:NB}

As discussed in section \ref{sec:naive-bayes}  \textit{Gaussian Naive Bayes} is parametrized by the following variables: the class prior probabilities $P(Y=c) = \frac{N_c}{N}$ , $\mu_{jc} $ and $ \sigma^2_{jc} $ the parameters explaining the conditional density distribution. These parameters can be computed as shown below
\begin{displaymath} N_c = \sum_{i=1}^n [Y^{(j)}=c] \end{displaymath}
\begin{displaymath} \mu_{jc} = \frac{\sum_{i=1}^n x_j^{(i)} [Y^{(j)}=c]}{N_c} \end{displaymath}
\begin{displaymath} \sigma_{jc}^2 = \frac{\sum_{i=1}^n (x_j^{(i)}[Y^{(j)}=c])^2}{N_c}-\left(\frac{\sum_{i=1}^n  x_j^{(i)}[Y^{(j)}=c]}{N_c}\right)^2  \end{displaymath}

We maintain $N_c$ for each class in the data set, which is the number of samples belonging to each class. In order to calculate $\mu_{jc}$ we maintain the sum of feature $j$ over the samples in class $c$, represented by $S_{jc}$. Similarly for $\sigma_{jc}$ we maintain the sum of squares of the values of feature $j$ in class $c$, represented by $SS_{jc}$. Maintaining the statistics above we calculate all the parameters of the model. Assuming we have $C$ classes in total in the data set, we need to maintain $O(d\times C)$ values. This is again independent of the number of training examples ($n$). 

The multinomial Naive Bayes model also has the same class prior probabilities $P(Y=c) = \frac{N_c}{N}$. In addition we have to maintain $ \theta_{ci} $ for which we need to also store $N_{ci}$ and $N_c$. These parameters are expressed as sum of feature values across the classes. For the case of the multinomial model, we need to maintain $O(d \times C)$ number of parameters for the model.

\subsection{Incremental Model Updates}
\label{subsec:incr_model}

In this section we demonstrate how incremental changes (data additions or deletions) can be supported by
the two models considered. Formally, let $M$ be a model build on data set $D$ consisting of points $n$.  We will demonstrate the incremental changes by considering adding point $(p_1,...,p_d,y)$ to the data set $D$, where $d$ is the dimension of the data. We wish to find the parameters of the new model $M'$ for data set $D' = D \cup (p_1 \ldots p_d,y)$ of size  $n+1$. 

\subsubsection{Linear Regression}
For the linear regression model $M$ we have already computed matrix $A$ and $B$ on data set $D$. We will calculate the $A'$ and $B'$ on $D'$ by operating on $A$ and $B$ and updating them to reflect the new point. The equations below show how to update matrix $A$ and $B$:

\begin{displaymath} A'_{ij} =  \sum_{j=1}^n x_i^{(j)} x_j^{(j)} + p_ip_j  \end{displaymath}
\begin{displaymath} B'_{i1} =  \sum_{j=1}^n x_i^{(j)} y^{(j)} + p_iy  \end{displaymath}

Deletions are handled similarly. Larger collections of points can be added/deleted in a similar fashion. Other statistics computed while building regression models like ANOVA table, AIC etc. which explain the goodness of fit of the model can also be incrementally maintained in a similar fashion. Details have been omitted for brevity. 

\subsubsection{Naive Bayes Classifier}

For the Naive Bayes model $M$ we have computed $N_c$ , $S_{jc}$ and $SS_{jc}$ on $D$. We can update these statistics for $D'$ according to the equations below

\begin{displaymath} N'_c = N_c  + [y=c]\end{displaymath}
\begin{displaymath} S'_{jc} = S_{jc} + p_j[y=c] \end{displaymath}
\begin{displaymath} SS'_{jc} = SS_{jc} + p_j^2[y=c] \end{displaymath}

Given that we have the updated statistics we can compute the parameters of the updated model $M'$. 
Similar observations hold for deleting data as well as operating on collections of points.

\subsection{Combining Models}
\label{subsec:lrm}

Let $D$ be the underlying data set of $n$ points. Assume that points in $D$ are associated with a unique
identifier, namely a point $p \in D$ is represented as $p = (id,y,\bf{x})$, where $id$ is the identifier, $y$ the dependent (class) variable and $\bf{x}$ the feature vector as before. To simplify notation for the remainder of the paper, we assume, without loss of generality that the unique identifier imposes a natural ordering in $D$. For example $id$ could be a time-stamp associated with the point (indicating the time it was generated). 
Casting our entire framework for the case where the points of the underlying data set $D$ do not have
a unique ordering is indeed possible. It requires however a different methodology and we defer description of
this case in our subsequent future work.
Also for brevity we will denote as $D_i$ both the model and the data set (subset of D) for which we wish to build a model on. A sequence of these data point identifiers determines a {\em model descriptor} which is a range of points in $D$. Let $D_1$ and $D_2$ be data sets represented by model descriptors $d(D_1)= [a_1,b_1]$ and $d(D_2) = [a_2,b_2]$. Our aim is to compute the model $D_c = D_1 \cup D_2$

We discuss the linear regression case. Naive Bayes models are handled similarly so we omit the description for brevity. Let $D_1$ and $D_2$ be two linear regression models. For each model we maintain the associated matrices $A = X^TX$ and $B=X^Ty$ along with the model descriptor signifying the data set on which it was calculated. 
Computing the regression model  $D_c = D_1 \cup D_2$ , involves considering two cases:
\textit{Case 1}: The two data sets do not have any points in common i.e. $ D_1 \cap D_2 = \phi$; this case can be easily identified by comparing the model descriptors of the two data sets. A specific entry in the matrix $X^TX$ for model $D_1$ looks like $\sum\limits_{j}^{D_1} x_a^{(j)} x_b^{(j)}$, where $a$ and $b$ 
are any two features. Thus, it can be seen that the corresponding matrix $A$ on data set $D_c$ can be computed as
\begin{displaymath} \sum\limits_{j}^{D_c} x_a^{(j)} x_b^{(j)} = \sum\limits_{j}^{D_1} x_a^{(j)} x_b^{(j)} + \sum\limits_{j}^{D_2} x_a^{(j)} x_b^{(j)}\end{displaymath}
which is essentially adding the corresponding elements of matrix $A$ of the two models directly.
 
\textit{Case 2}: The two data sets have points in common i.e $ D_1 \cap D_2 \neq \phi$; in this case the points common to both data sets can be determined from the corresponding model descriptors. If we directly operate on the two models the points which are common will be accounted for twice. Thus, we need to exclude points represented in both model and make sure we account for them once in the final model.
We compute matrix $A$ on  data set $D_c$ as follows:
\begin{displaymath} \sum\limits_{j}^{D_c} x_a^{(j)} x_b^{(j)} = \sum\limits_{j}^{D_1} x_a^{(j)} x_b^{(j)} + \sum\limits_{j}^{D_2} x_a^{(j)} x_b^{(j)} - \sum\limits_{j}^{D_1 \cap D_2} x_a^{(j)} x_b^{(j)}\end{displaymath}
\begin{displaymath} \sum\limits_{j}^{D_c} x_a^{(j)} x_b^{(j)} = \sum\limits_{j}^{D_1} x_a^{(j)} x_b^{(j)} + \sum\limits_{j}^{D_2 - D_1} x_a^{(j)} x_b^{(j)}\end{displaymath}
\begin{displaymath} \sum\limits_{j}^{D_c} x_a^{(j)} x_b^{(j)} = \sum\limits_{j}^{D_2} x_a^{(j)} x_b^{(j)} + \sum\limits_{j}^{D_1 - D_2} x_a^{(j)} x_b^{(j)}\end{displaymath}

The matrix $X^Ty$ for $D_c$ can be computed in a similar fashion. Notice that in this case we need to retrieve a few extra points from $D_1,D_2$. This incurs an IO cost that needs to be accounted for (see section \ref{sec:opt}).
 
\end{sloppypar}

\section{Incremental Logistic Regression Models} \label{sec:logreg}

\label{sec:logistic}
Stochastic Gradient Descent(SGD) is a popular optimization framework for estimating parameters of a Logistic Regression model. 
SGD is a sequential algorithm that updates weight parameters at each iteration until convergence. A typical drawback of SGD is its poor scalability on large data sets. Recognizing the importance of analytical tasks on massive data sets, recent work has established methodologies to scale SGD into realistic data sets \cite{Mann:SGD,Zinkevich:PSGD}. We adopt such methodologies and extend them to
fit our framework.

A generic loss function for the Logistic Regression model is given in Equation \ref{eqn:logistic}. SGD is applied to identify the model parameters $w$ which minimize the loss function. 
We describe a variant of the SGD algorithm called Mixture Weight Methods \cite{Mann:SGD}.
Let us consider a sample $S = (S_1,....,S_p)$ of $pm$ points formed by $p$ sub-samples of $m$ points each drawn i.i.d, $S_1,...S_p$.
Algorithm \ref {algo:psgd} outlines the steps for executing Mixture Weight Method. Notice that the outer-loop of the algorithm can be executed in parallel and as a result the approach can easily utilize multiple processors if required. 

\begin{algorithm} 
\caption{Mixture Weight Method} \label{algo:psgd}
\begin{algorithmic}[1]
\For { $\bf{all}$ $i \in \{1,...p\}$}
\State $\vec{w_i} \leftarrow $ 0 
\For { $t \leftarrow$ 1 to $T $}
\State $\nabla F_{S_i}(w) \leftarrow $ GRADIENT($F_{S_i}(w)$) 
\State $w_i \leftarrow w_i + \lambda(\nabla F_{S_i}(w)) $
\EndFor
\EndFor
\State Aggregate all $w_{\mu} = \sum\limits_{k=1}^{p}\mu_k w_k$
\end{algorithmic}
\end{algorithm}

Where $F_{S_i}$ is the optimization function for sample $S_i$ and $T$ is the number of iteration required to converge. Thus, algorithm \ref{algo:psgd} computes the model parameters on subsets of data and then averages the parameters across all the subsets to compute the parameter for the complete set of data. In \cite{Mann:SGD} it is shown that Algorithm \ref{algo:psgd} has good convergence properties and under certain assumptions establishes a relationship between the $w_{\mu}$ estimated and the values computed executing SGD on the entire data set.

We extend this idea in our framework as well. Let $D$ be an underlying data-set of size $n$ and a point $p \in D$ is represented as $p = (id,y,\bf{x})$, where $id$ is the identifier, 
$y$ the dependent (class) variable and $\bf{x}$ the feature vector as before. 

A request to create a logistic regression model on data set $D_q$ (the query set), is represented by a range of $id$ values $[a,b]$ over $D$ such that $b-a = |D_q|+1$. The query data set is segmented into smaller chunks of equal size $l$ with the obvious assumption that $l\leq |D_q|/2$. This results into $\lfloor \frac{|D_q|}{l} \rfloor$ number of chunks of equal size. These chunks are created in the increasing order of ID values. A chunk $S_i$ is given by the following range
\begin{displaymath} S_i = [a + (i-1)*l ,  a + i*l ]\end{displaymath}
and $ i \in \{1, ... ,\lfloor \frac{|D_q|}{l}  \rfloor \} $. Assuming that the logistic regression models for each chunk are available, they are combined in the spirit of algorithm \ref{algo:psgd} and produce the model for $D_q$. Assuming that none of the chunks is available, a request to build the model for $D_q$ can utilize the base data to build the logistic regression model. At the same time, the chunks are generated for $D_q$, the logistic regression model build for each of them, and the result is materialized in order to benefit future model creation requests.

Any request to build a logistic regression model for a data set $D^{'}_q$ first tests whether $D^{'}_q$ contains 
any of the chunks for which a model has already been materialized. If it does we can readily
utilize its parameters and save computation time. Any parts of $D^{'}_q$ that are not currently "covered" by existing chunks
have to be computed from the base data set. Thus, we retrieve the parts of $D^{'}_q$ for which we don't have the model,
generate chunks of size $l$ and compute the model parameters for them. Finally we average all parameters from all chunks to compute the model.  Algorithm \ref{algo:chunkalg} presents our overall approach. 


\begin{algorithm} 
\caption{Incremental Logistic Regression} \label{algo:chunkalg}
\begin{algorithmic}[1]
\Procedure {Incremental Logistic Regression}{$D_q$}
\State $S$ $\leftarrow$ ranges in $D_q$ for which a model already exists
\State $P$ $\leftarrow$ \{\}
\For {all the ranges $r \in S$}
\State $D_q \leftarrow D_q - r$
\State $P_i \leftarrow$ Linear Regression parameters for r
\State $P$ $\leftarrow$ $P \cup P_i$
\EndFor
\State Sort $D_q$ in increasing order of $ID$ values
\State Create chunks of size $l$ from $D_q$
\State Compute Linear Regression parameters on each chunk $l$ and add to $P$
\State Average all parameters in $P$
\EndProcedure
\end{algorithmic}
\end{algorithm}

Theorem \ref{theorem:avg_bound} establishes a relationship between the outcome of Algorithm \ref{algo:chunkalg} on $D^{'}_q$ and that computed by applying SGD directly on $D^{'}_q$.






\begin{theorem}\label{theorem:avg_bound}
Let $w_{\mu} $ denote the mixture of weight vector obtained by applying Algorithm \ref{algo:chunkalg} on a  model query $D_q$ and $\mu_{SGD}$ be the weight vector computed by applying SGD on $D_q$. Then, for any $\delta >0$, with probability at least $1 - \delta$, the following inequality holds:
\begin{displaymath} 
 \| w_{\mu} - w_{SGD} \| \leq \frac{R\sqrt{2}}{\lambda}(\frac{1}{\sqrt{l}}+\frac{1}{\sqrt{|D_q|}}) + \frac{2\sqrt{2} R}{\lambda\sqrt{pl}}\sqrt{log1/\delta} \end{displaymath}
\end{theorem}

where $R$ is the bound for the norm of feature vectors, $\lambda$ is the regularization constant, $p = \lfloor \frac{|D_q|}{l} \rfloor$ is the number of chunks of $D_q$ created in step 10 of Algorithm \ref{algo:chunkalg}, $l$ is the size of each chunk and $1-\delta$ represents the probability with which this inequality holds. The proof of \ref{theorem:avg_bound} follows the methodology presented in \cite{Mann:SGD} and is
available in the full version of the paper \cite{full:paper}.


Note that in contrast to the discussion of section \ref{subsec:incr_model}, for logistic regression models, this framework supports adding points to an existing model not deleting them. Thus we can construct new models only by adding points to existing models (combining existing chunks). This is inherent to the nature of the approximation of the logistic regression. As a result the space of all possible options to consider when creating a new model considers addition of points to an existing model, not deletions.

\section{Optimization Considerations} \label{sec:opt}

Given a collection of materialized models over a data set $D$, it is evident that a request to create a new model $D_q$ can
readily utilize existing models. We seek to understand the trade offs involved while building the new model $D_q$. Several
options are available including building $D_q$ by manipulating data from $D$ or utilizing materialized models directly
and/or suitably adjusting them using data from $D$. 

\begin{figure*}
    \centering
    \begin{subfigure}[b]{0.35\textwidth}
        \centering
        \raisebox{0.4in}{\includegraphics[width=2.4in,height=1.2in]{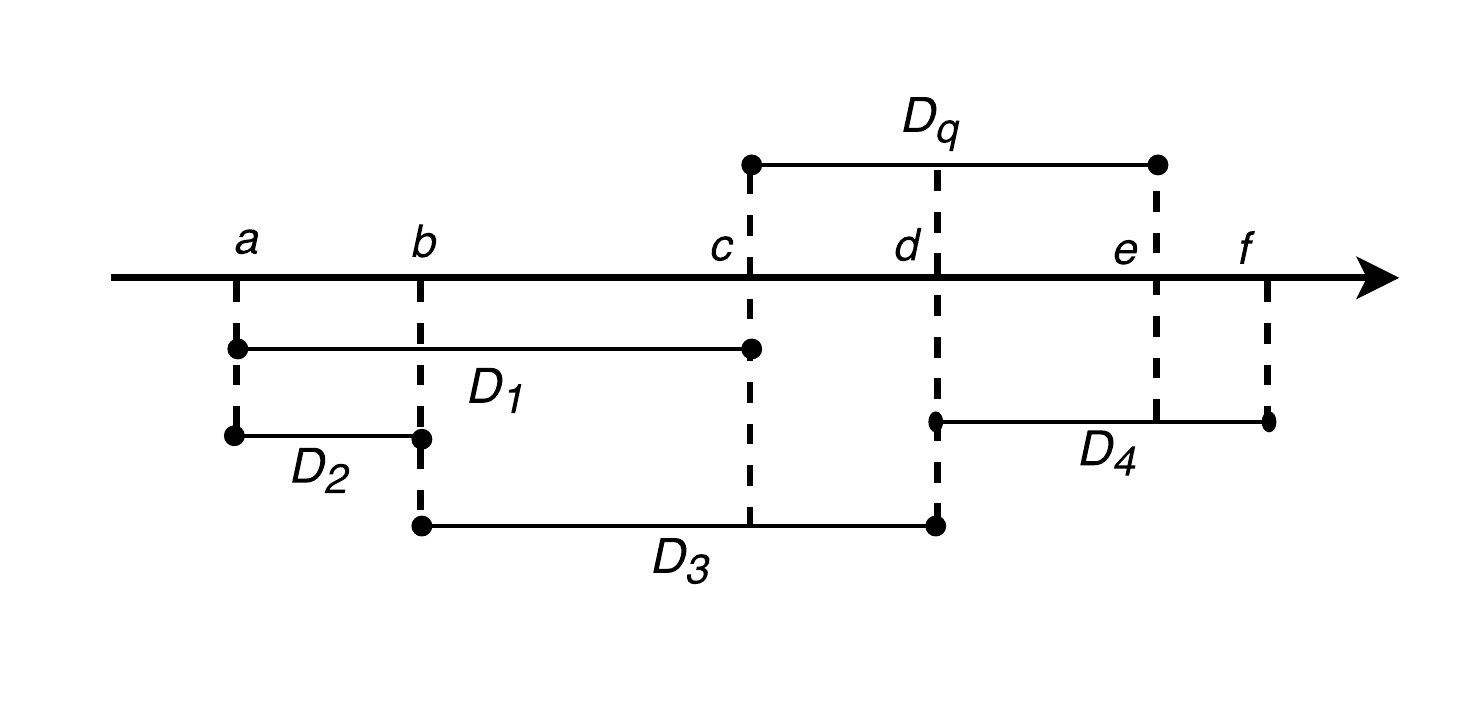}}
        \caption{Materialized model state}
        \label{figure:system-state}
    \end{subfigure}%
   ~
    \begin{subfigure}[b]{0.25\textwidth}
        \centering
        \raisebox{0.4in}{\includegraphics[height=1.2in]{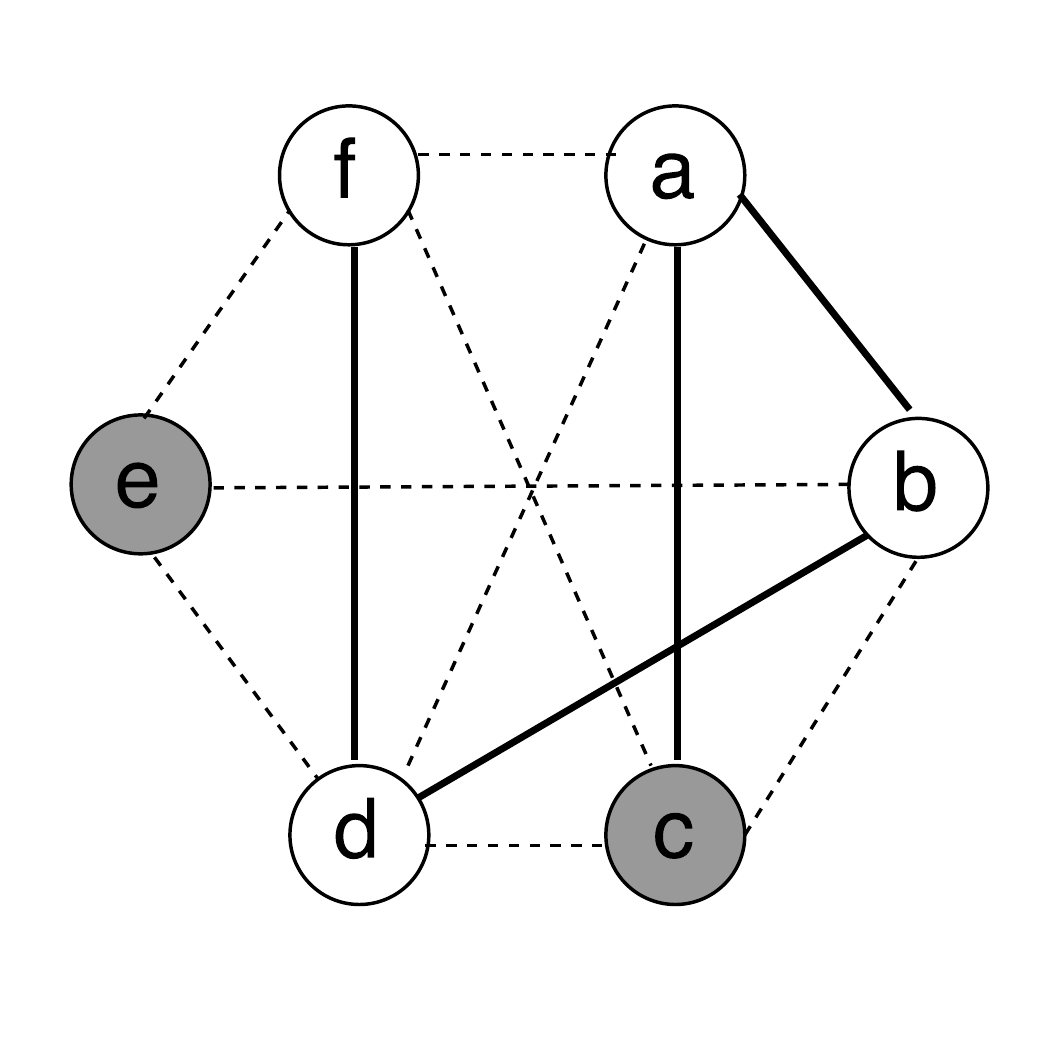}}
        \caption{Query Graph using cost model}
        \label{figure:query-graph}
    \end{subfigure}%
    ~
    \begin{subfigure}[b]{0.35\textwidth}
        \centering
        \includegraphics[width=2in,height=2in]{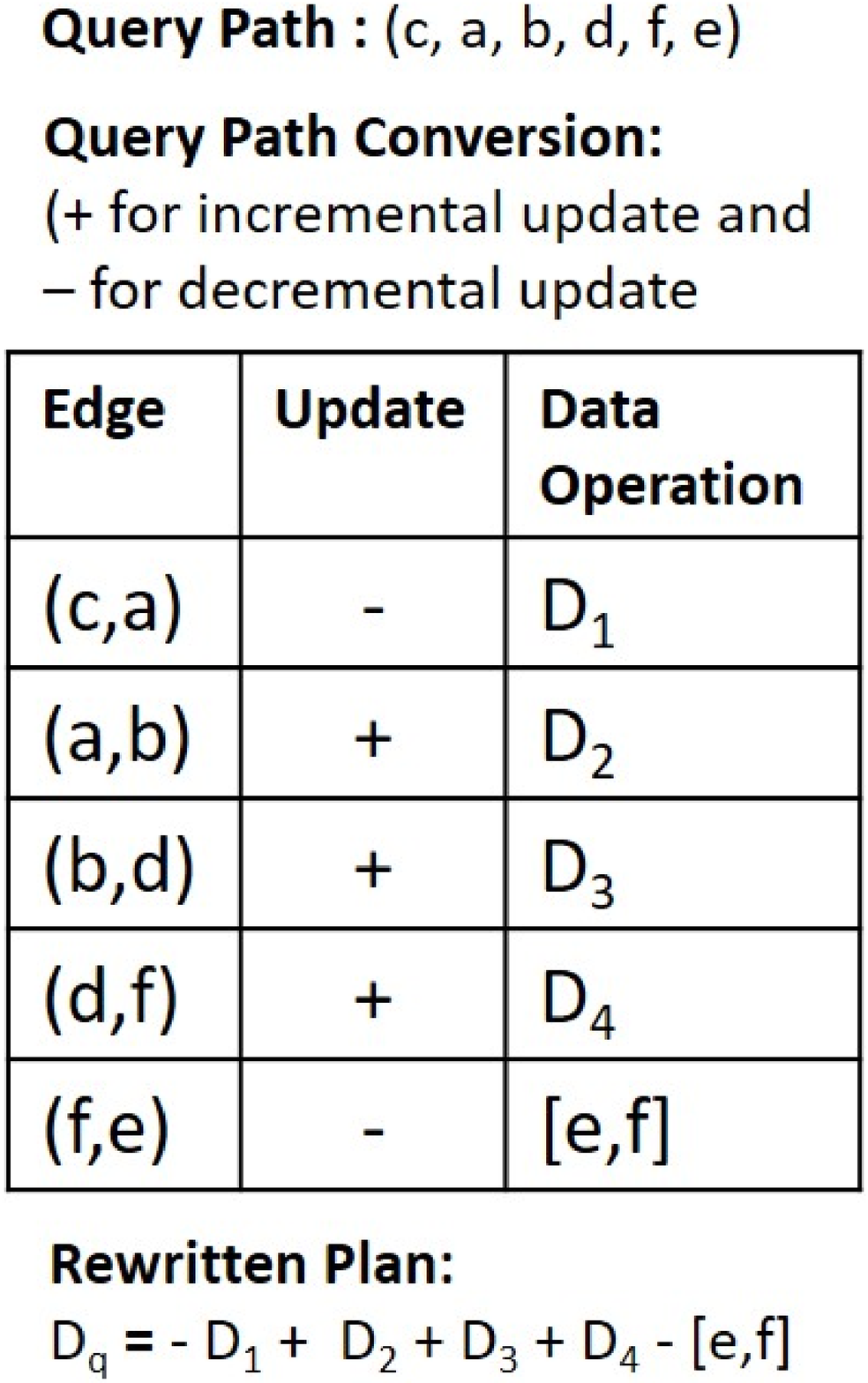}
        \caption{Query Path Conversion}
        \label{figure:plan-conversion}
    \end{subfigure}
    \caption{Graph modeling to find optimal execution strategy for query interval $I_Q$}
\end{figure*}

Consider Figure \ref{figure:system-state}. It depicts data set $D$ and four materialized models ($D_1,D_2,D_3,D_4$). A
request to build model $D_q$ is faced with numerous options. 
Using the materialized models to generate model $D_q$ , Equations \ref{equation:interval-decomposition1}, \ref{equation:interval-decomposition2} and \ref{equation:interval-decomposition3}  show different ways in which this can be achieved 

\begin{align}   \label{equation:interval-decomposition1}
    D_{q} = D_{3} + D_{4} - [b,c] - [e, f]. 
\end{align}
\begin{align}    \label{equation:interval-decomposition2}
    D_{q} = D_{3} + D_{4} - (D_{1} - D_{2}) - [e, f]. 
\end{align}
\begin{align}   \label{equation:interval-decomposition3}
    D_{q} = [c,d] + D_{4} - [e, f]. 
\end{align}

Equation \ref{equation:interval-decomposition1} represents an execution strategy which will fetch models $D_{3}$ and $D_{4}$ combine them, then remove all points in the range of $[b,c]$ and $[e, f]$ (this constitutes incrementally updating, removing these points, from the combined model). This step consists of accessing $D$ and retrieving all points between $[b,c]$ and $[e,f]$. In equation 
 \ref{equation:interval-decomposition2} instead of retrieving $[b,c]$ from $D$, we compute that operation by manipulating (subtracting) models $D_2$ and $D_1$. If the model allows (e.g., linear regression) we can subtract $D_2$ from $D_1$ and compute the model for $[b,c]$ directly.
Similarly, Equation  \ref{equation:interval-decomposition3} 
represents another execution strategy which involves retrieving $D_4$ along with data points between $[c,d]$ and $[e,f]$ 
and manipulating them (incrementally updating, adding and removing points) to complete the model construction. Other choices are also possible including retrieving all points
between $[c,e]$ from $D$ and computing the model directly from base data. In order to be able to quantify the merits of each choice, as is typical in cost based query optimization \cite{Chaudhuri:QueryOpt} we need to a) assess all possible choices efficiently and b) quantify the cost of each option in order
to determine the least cost way to build the model.

The specifics of the cost model are orthogonal to our approach. The cost depends on the type of model and also the model descriptor which
may or may not involve disk access. In addition retrieving data from $D$ typically involves disk access. The only requirement we impose in the cost model adopted is to be monotonic. This means that all things being equal, the cost of retrieving a certain number of data points from disk should be at least as costly as the cost of retrieving less points. For the remainder of the paper we assume a cost model $C$ that
is monotonic. To facilitate notation the cost of using a materialized model $D_i$ is denoted as $C(D_i)$. The cost of retrieving $n$ data points from disk is denoted as $F(n)$.

Let $S$ be a collection of materialized models on data set $D$. For a model $D_q$, let $d(D_q) = [l_q,u_q] $ be a model descriptor on which a new model 
has to be computed. $l_q$ and $u_q$ in this case express a range of data points on $D$. 
We wish to identify the minimum cost collection of materialized models and/or data points from $D$ that would be used to construct the model for $d(D_q)$, $D_q$.

\newdef{definition}{Definition}
\begin{definition} 
Let $d(D_q) = [l_q,u_q] $ represent a model descriptor for model $D_q$ which we wish to construct and $S$ be the set of available materialized models. Then the set $S_R \subseteq S $ of {\em relevant models} for $D_q$ is defined as follows :
\begin{enumerate}
\item If for a materialized model $S_i \in S$, $d(S_i) \cap q \neq \varnothing$, then $S_i \in S_R$.    
\item $\forall S_i' \in S$ such that $\exists S_j \in S_R$ with $d(S_i') \cap d(S_j) \neq \varnothing$ then $S_i' \in S_R$.
\end{enumerate}
\end{definition}

Intuitively the models in $S_R$ are {\em relevant models} because they either contain common data points with the ones of interest to $D_q$ and/or they are models that can be manipulated (by combinations of models or incremental updates of models) to produce models that assist in computing $D_q$. As we can see in Figure \ref{figure:system-state} materialized models $D_3$, $D_4$ contain data points common with $D_q$ while $D_1$ and $D_2$ can be manipulated along with $D_3$ to produce models relevant to the computation of $D_q$. While computing
$D_q$, only relevant models will be part of $S_R$.

\begin{algorithm}  
\caption{PreprocessDescriptors $(S)$} \label{algo:merge}
\begin{algorithmic}[1]
\State \emph{enhancedDescriptors} $\leftarrow$ mapping of descriptors and the corresponding materialized models 
\State \emph{descriptor} $\leftarrow$ a model descriptor represented by $[l,u]$
\State \emph{arrayDescriptors} $\leftarrow$ array of descriptors
\State Sort S in increasing order of $l$ values
\State \emph{descriptor}[0] $\leftarrow$  $l$ value of first descriptor in S
\State \emph{descriptor}[1] $\leftarrow$  $u$ value of first descriptor in S
\State arrayDescriptors $\leftarrow$ append first descriptor in S 
\For { each descriptor $r \in S$ }
\If{ $r$ overlaps \emph{descriptor}}
\State \emph{descriptor}[1] $\leftarrow$ max(\emph{descriptor}[1], $u$ value of $r$)
\State \emph{arrayDescriptors} $\leftarrow$ append $r$
\Else{}
\State enhancedDescriptors.put(\emph{descriptor},\emph{arrayDescriptors})
\State arrayDescriptors $\leftarrow$ \{\}
\State arrayDescriptors $\leftarrow$ append $r$
\State \emph{descriptor}[0] $\leftarrow$  $l$ value of $r$
\State \emph{descriptor}[1] $\leftarrow$  $u$ value of $r$
\EndIf
\EndFor
\State return \emph{enhancedDescriptors}
\end{algorithmic}
\end{algorithm}

The set of relevant models $S_R$ is important since it accurately reflects the set of models to be considered during the computation of
$D_q$. Instead of assessing all relevant models every time a new request for a model $D_q$ arises, we pre-process the collection of all materialized models $S$ to facilitate the derivation of $S_R$ for a given $D_q$. Thus given $S$ we pre-process it to facilitate the computation of relevant models. Algorithm \ref{algo:merge} presents the overall approach. The basic idea is to pre-process $S$ and create {\em enhanced} descriptors that are the union of multiple model descriptors. Such enhanced descriptors can facilitate quick search for relevant models.

Running algorithm \ref{algo:merge} in the example of Figure \ref{figure:system-state} will produce two
enhanced descriptors namely $[a,d]$ formed by combining descriptors for models $\{ D_1, D_2, D_3 \}$ and $[d,f]$ which constitutes the descriptor of model $\{D_4\}$.

Maintaining $enhancedDescriptors$ makes it easier to compute the set $S_R$. When the descriptor of a model $D_q$ is provided, we compare it against the $enhancedDescriptors$. If a descriptor intersects any of the descriptors in $enhancedDescriptors$ all the materialized models mapped to that descriptor become part of $S_R$.

Algorithm \ref{algo:merge} will produce the set $S_R$ of all models that should be considered in deriving
model $D_q$. Using the descriptors in $S_R$ we create a complete undirected graph $G(V,E)$ where each node $v \in V$ corresponds to the $l$ or $u$ values of the model descriptions in $S_R$. 
As for our running example the set $S_R$ contains models $D_1$ to $D_4$. Thus we add the $l$ and $u$ values of the descriptors of these materialized models. As we can see in figure \ref{figure:query-graph} it contains 
$a$ to $f$ as nodes. An edge $\epsilon \in E$ corresponds to the cost of building a model for the data set specified by the two nodes adjacent to $\epsilon$. If materialized model $M$ exists for the data descriptor specified by the nodes adjacent to the edge $\epsilon$ then the cost of the edge is the cost of using model $M, C(M)$. If a model does not exist for that data set the cost of that edge is determined by the number of points in the range. In our example the solid edges in our graph represent the materialized models $D_1$ to $D_4$. For all the other edges the cost is given by $F(n)$, where $n$ is the number of points in the interval represented by the edge. 
Given $D_q$ and $d(D_q)=[l_q,u_q]$ values $l_q,u_q$ represent the source and destination respectively. These are shown as grey nodes in Figure \ref{figure:query-graph}.

\begin{algorithm}[!htb] 
\caption{Identify Optimal Execution Path} \label{algo:optimization}
\begin{algorithmic}[1]
\Procedure {GenerateGraph}{$S_R,D_q,C(M),F(n)$}
\State initialize Graph $G(V,E)$
\For {each descriptor $r \in S_R$}
\State G $\leftarrow$ add vertices corresponding to $l$ and $u$ values of $r$
\State G $\leftarrow$ add an edge between two new vertices with weight $C(D_r)$
\EndFor
\For {each vertex $v \in G$}
\For {each vertex $u \in G$}
\If {(no edge between $u$ and $v$) \& $u\neq v$ }
\State G $\leftarrow$ add an edge b/w $u$ \& $v$ with weight $F(|u-v|)$
\EndIf
\EndFor
\EndFor
\State return $G(u,v)$
\EndProcedure
\Procedure {OptimalPath}{$S_R,D_q,C(M),F(n)$}
\State Identify $S_R$ using algorithm PreprocessDescriptors
\State G $\leftarrow$ GenerateGraph($S_R,D_q,C(M),F(n)$)
\State Apply Dijkstra's Algorithm using $d(D_q)$ $l$ and $u$ values as source/destination
\State Return the shortest path 
\EndProcedure
\end{algorithmic}
\end{algorithm}

Every path from source node to destination represents an execution strategy to construct model $D_q$. Figure \ref{figure:plan-conversion} illustrates how to convert a path on the
graph to a set of operations that compute the model. Consider a path on the graph represented by the following sequence of nodes $(c,a,b,d,f,e)$. We fetch four materialized models $D_1, D_2, D_3 $ and $D_4$ for the 
edges $(c,a), (a,b), (b,d) $ and $(d,f)$ respectively. The edge $(f,e)$ does not correspond to any materialized model , thus cost of that edge is equivalent to fetching the corresponding data points from disk. The decision 
whether to manipulate an existing model by adding or removing data points from it is decided by the nodes of the edge. If we traverse the edge $(i,j)$ from $i$ to $j$ and $i > j$ then we remove points from the model  
otherwise we add data points. In our example edge $(c,a)$ $c > a$ (as indicated in Figure \ref{figure:system-state}) and that constitutes removing points. The total cost of a query path is given by 
\begin{displaymath}
C(D_q) = \sum\limits_{i}^{k} cost(e_i) + (k-1) * c_{merge}
\end{displaymath}
where $cost(e_i)$ is cost of each edge and $c_{merge}$ is cost of merging two materialized models. The cost $c_{merge}$ depends on the type of model under consideration. For example for linear regression the cost is outlined in section \ref{subsec:lrm}. It involves (after retrieving the model parameters) a simple manipulation of corresponding model representations. It is expected that the cost of merging two materialized models is much 
less than the cost of fetching models or the cost of fetching  data points from the disk $(c_{merge} \ll e_i)$.
Depending on how the model descriptors and model parameters are stored, retrieving them may not require any disk access. For example in the case of a linear regression model, the model descriptors would be just a range of values and the model parameters would be as outlined in Section \ref{subsec:lr}.

\begin{figure*} 
    
    \begin{subfigure}[b]{0.25\linewidth}
        \centering
        \includegraphics[width=2.2in,height=1.5in]{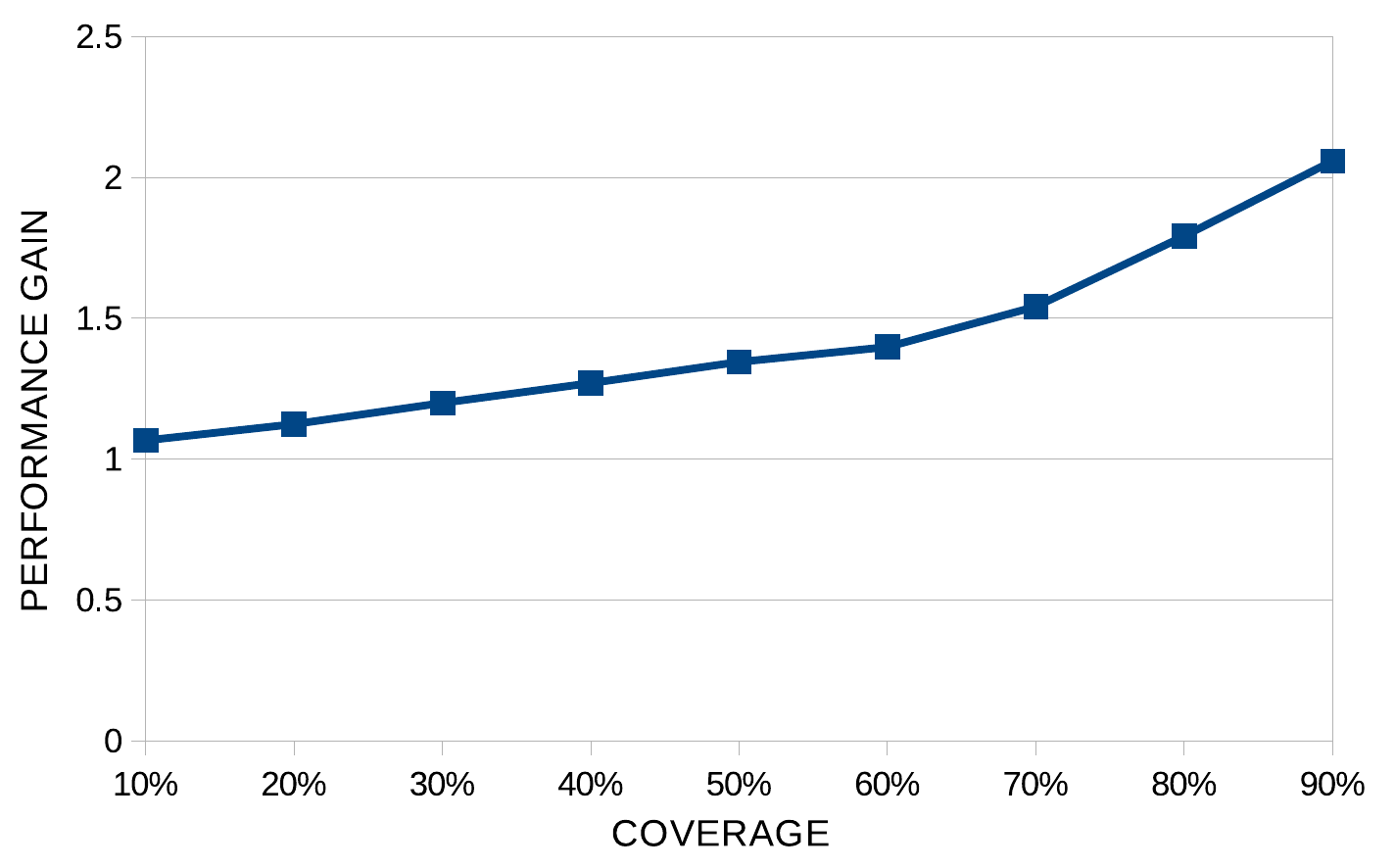}
        \caption{Naive Bayes Model}
        \label{figure:NB_speed_up}
    \end{subfigure}%
   \hspace{1.5cm}
    \begin{subfigure}[b]{0.25\linewidth}
        \centering
        \includegraphics[width=2.2in,height=1.5in]{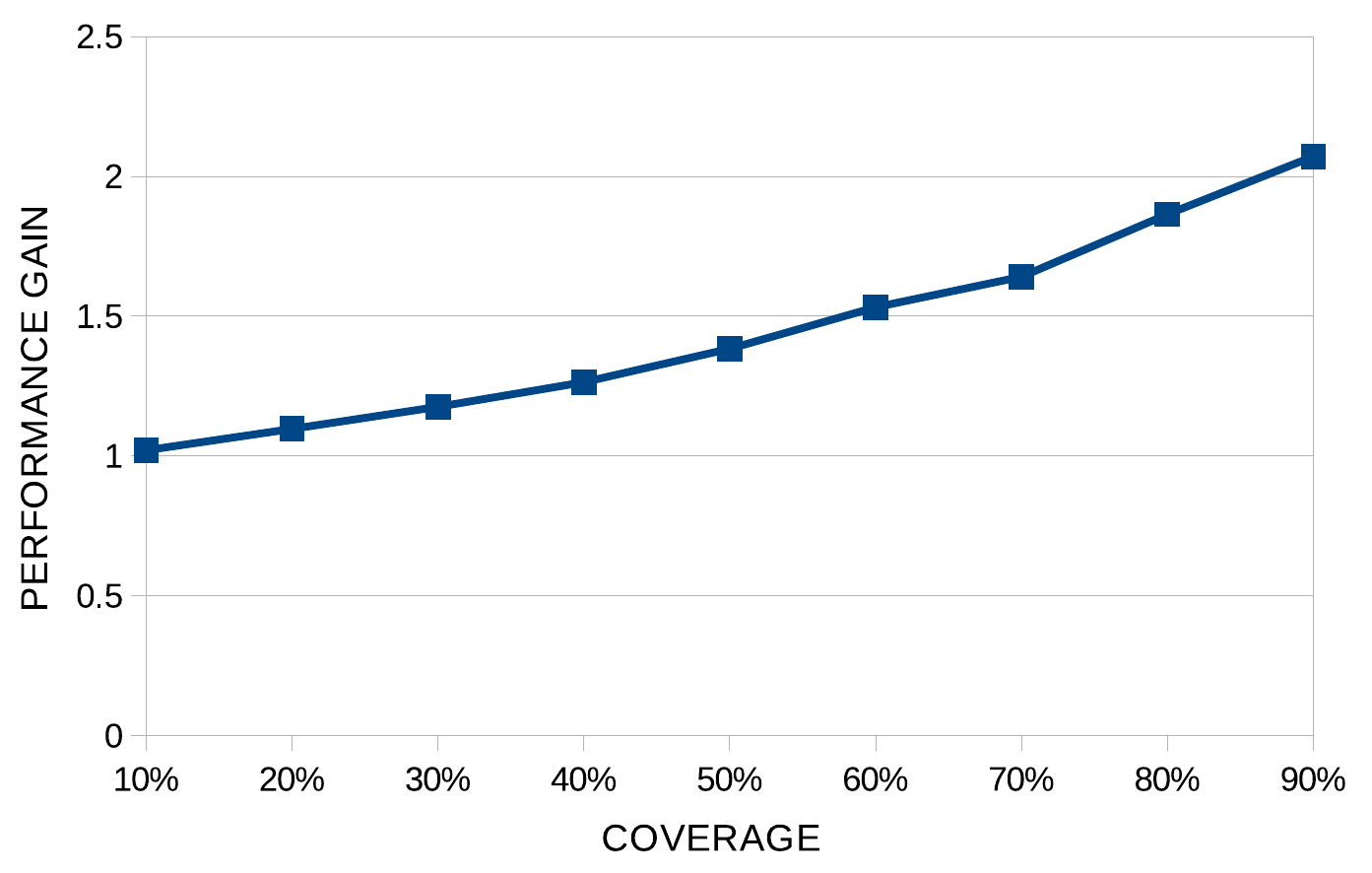}
        \caption{Linear Regression Model}
        \label{figure:LR_speed_up}
    \end{subfigure}%
   \hspace{1.5cm}
    \begin{subfigure}[b]{0.25\linewidth}
        \centering
        \includegraphics[width=2.2in,height=1.5in]{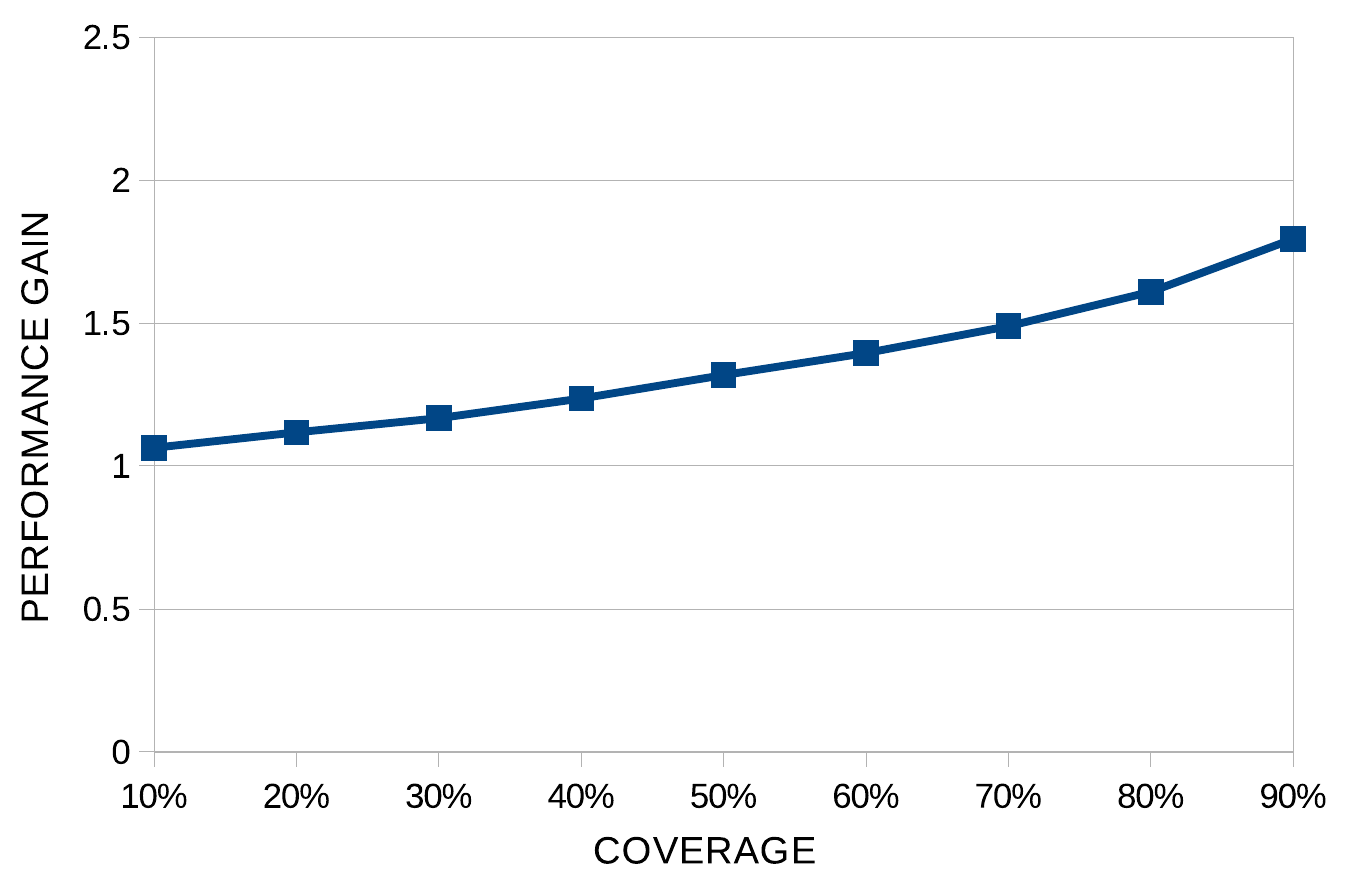}
        \caption{Logistic Regression}
        \label{figure:LogR_speed_up}
    \end{subfigure}
    \caption{Performance gain against coverage percentage }
    \label{figure:performance}
  
    \begin{subfigure}[b]{0.25\linewidth}
        \centering
        \includegraphics[width=3.75in,height=1.5in]{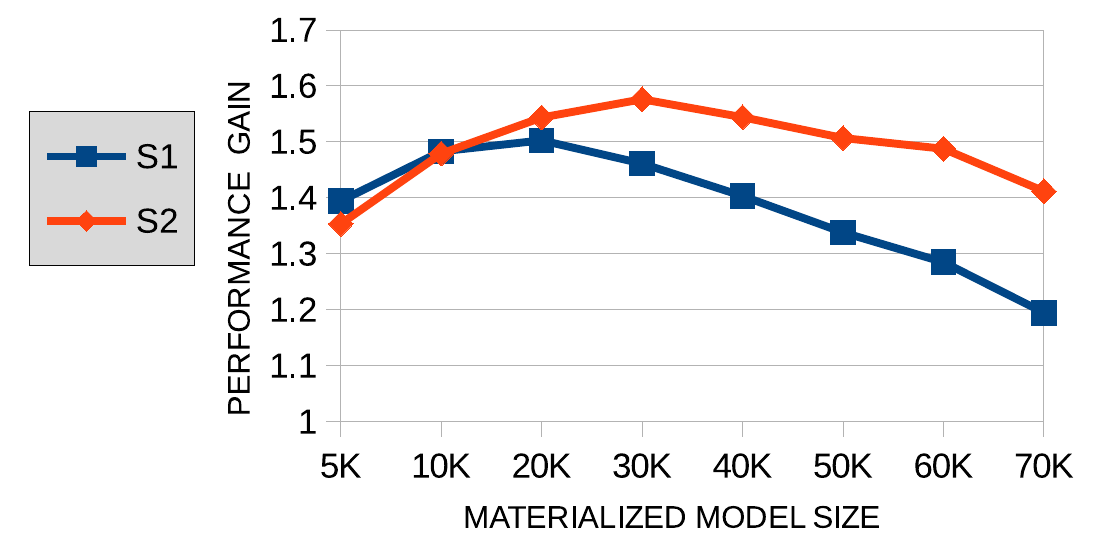}
        \caption{Naive Bayes Model}
        \label{figure:NB_summary_vs_speedup}
    \end{subfigure}%
   \hspace{5cm}
    \begin{subfigure}[b]{0.25\linewidth}
        \centering
        \includegraphics[width=3in,height=1.5in]{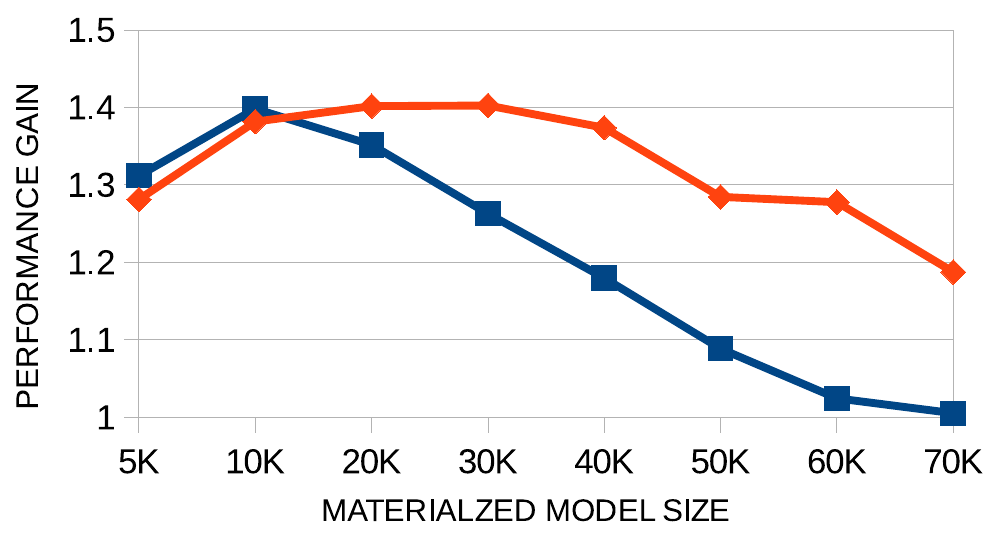}
        \caption{Logistic Regression Model}
        \label{figure:LogR_summary_vs_speedup}
    \end{subfigure}%
    \caption{Performance gain against materialized model size }
    \label{figure:summary_vs_speedup}
 
 \end{figure*}

It is evident that by construction the problem of identifying the minimum cost to construct the model $D_q$ is equivalent to identifying the shortest path from a single source in a weighted graph. Dijkstra's algorithm can be used to identify the optimal
solution in  $O(|E|log |V|)$, $|E|$ is the number of edges and $|V|$ is the number of vertices in the graph.

We presented the entire solution for the case of models that support addition and removal of points  to derive new models, as is the case of models such as linear regression and Naive Bayes. For the case of logistic regression removal of points is not supported in the model we utilize to approximate the regression. 
In this case we have to modify slightly the algorithm to enable optimization of logistic regression models as well. The changes are as follows:

\begin{itemize}
\item During identification of the set $S_R$ we will include models such that their descriptors are fully contained in the descriptor $d(D_q)$.
\item The graph $G$ constructed will only contain directed edges from nodes $i$ to $j$ such that $ i < j $. 
\end{itemize}

These two changes will enable algorithm \ref{algo:optimization} to operate on logistic regression models and yield the least cost options to construct such models as well.

\section{Experiments} \label{sec:exp}
%
%

In this section we present a detailed performance comparison of our entire approach and proposal compared to alternate approaches.
We utilize materialized models to save processing costs, while building new models for an incoming (model construction) query $D_q$ as described in section \ref{sec:opt}. The natural alternative is not to materialize models, but instead build the new model directly from the raw data. We compare our approach against this baseline.  Our aim from these experiments is three-fold : (a) Highlight
the factors that affect performance for our materialization framework and associated trade-offs. (b) Detail the impact of our optimization framework in terms of its overheads and benefits and (c) analyze the accuracy of logistic regression materialization framework. Note that for the case of the linear regression and naive Bayes models, the models we construct are exactly the same as those constructed by the baseline, so there are no accuracy trade offs in these cases.

\textit{\textbf{Data}}. We test our framework utilizing synthetically generated data. Two different data set are generated for regression and classification problem. The choice of synthetic data allows us to change various parameters during experimentation. In addition experiments are focused on performance while scaling the size of the model and performance does not depend on quality of data but is governed by the size and type of data. The data is generated using publicly available synthesizers \cite{scikit-learn}. A random noise and interdependency among features is added while synthesizing data to simulate real world scenarios. In this section we present results using data sets up to 5 millions points with 10 features in each point. We tested all algorithms with synthetically generated data sets of larger sizes but the trends observed in our experiments were nearly the same. In addition we utilized popular real data sets from UCI Machine learning repository \cite{UCI:MLR} in our experiments and in all cases the results are consistent with those presented herein for synthetic data sets. 

\textit{\textbf{Experimental Setup}}. All our experiments were prototyped on top of MySQL(version 5.5.44) in a single node RDBMS setting. The model materialization framework code has been written in Python. The experiments were carried out on a PC running Linux Kernel Version 3.13.0-43-generic. The machine has a 3.40GHz Intel Core i7-3770 CPU with 16 GB of main memory.

Our framework is naturally parametrized by the size of the materialized models $(l)$ and the size of the incoming model construction query $(D_q)$. Another important parameter which is implicit in our discussion is the amount of data covered by the materialized models. Materialized models can be spread uniformly across the data set or may be concentrated on a few data points. To quantify the coverage we compute the number of unique data points covered by the materialized models and express it as a percentage of the total size of the data set. Formally let $D_1,...,D_n$ be the collection of models materialized at a given stage in the framework. For the data set, $D$, coverage is defined as follows :

\begin{displaymath} Coverage(\%) = \frac{|D_1 \cup D_2 ... \cup D_n|}{|D|} \times 100
\end{displaymath}

These parameters are varied across our experiments to understand their impact on performance gain.  Let $D_q$ be a model construction query. Our optimization framework identifies the optimal way to build model $D_q$. Let the overall time taken by our framework to build the model be $T$ (including the optimization and model construction time). Let the time taken by the baseline be $T_0$. Then the performance gain is calculated as follows 

\begin{displaymath} Performance \ Gain (PG) = \frac{T}{T_0}
\end{displaymath}

  In all experiments we report expected numbers. A query set $S$ containing one thousand queries is generated for each experiment. The query size is chosen from a uniform or normal distribution as explained in individual sections. These queries can represent a range of data points which is positioned anywhere across the underlying data. Similarly the materialized model size $(l)$ is also chosen from a uniform distribution, normal distribution or a fixed size. We create a set of materialized models $M$ on the data set with a given coverage as required in the experimental setting. The models are materialized before executing the query set $S$.

\subsection{Analyzing Performance}

 We assess the overall performance gain attained by our approach as compared to the baseline. Experiments were run for all three machine leaning models Linear Regression, Naive Bayes and Logistic Regression. The sizes of the sets $M$ and $S$ are chosen from the same normal distribution, $\mathcal{N}(50K,12.5K)$.  The x-axis depicts the percentage of data covered by materialized models.
We execute the queries in set $S$ and report the performance gain. Figure \ref{figure:NB_speed_up} and \ref{figure:LR_speed_up}  show that we were able to achieve a performance gain of \textbf{2x} as the coverage reaches 90\%. The increase in coverage implies a higher probability of identifying  {\em relevant models} for the query. Thus the expected performance gain improves as the coverage increases. The performance gain for Logistic regression is shown in Figure \ref{figure:LogR_speed_up}. The maximum performance gain achieved in logistic regression is \textbf{1.8x} which is slightly lower than the earlier two models. This can be explained by the fact that for Logistic Regression our framework supports only incremental updates to materialized models (section \ref{sec:logreg}). Thus, it eliminates certain execution strategies which would have been faster in the presence of decremental updates.  

\begin{table}[h!t]
\centering
 \begin{tabular}{||c | c||} 
 \hline
 Coverage & Model Sizes (MB) \\ [0.5ex] 
 \hline\hline
 20\% &  1.5\\ 
 \hline
 40\% &  1.8\\
 \hline
 60\% &  2.5\\
 \hline
 80\% &  3.5\\
 \hline
 90\% &  4.5\\
 \hline
\end{tabular}
\caption{Disk space occupied by materialized models for various coverage(\%)}
\label{table:size}
\end{table}

The previous experiment demonstrates that utilizing materialized models can have a profound effect on performance when constructing new.
However materializing a model comes at a cost, namely that of storing the model descriptors as well as the model details (e.g., regression parameters and meta-data  in the case of linear regression as defined in section \ref{sec:reg}). 
Table \ref{table:size} depicts the space occupied by the materialized linear regression models for each value of coverage. The size of the materialized model is fixed at 5K points. The base data set size is 350MB containing 5M points with 10 features. As it is visible from the table, the overheads in storage imposed by the materialized models is around 1.2\% of the original data. 
Similar trends hold for the other models of interest in our study. It is evident that the minor storage overheads are heavily compensated in light of the performance benefits.

\begin{figure*}
      \begin{subfigure}[b]{0.4\linewidth}
        \centering
        \includegraphics[width=2.5in,height=1.5in]{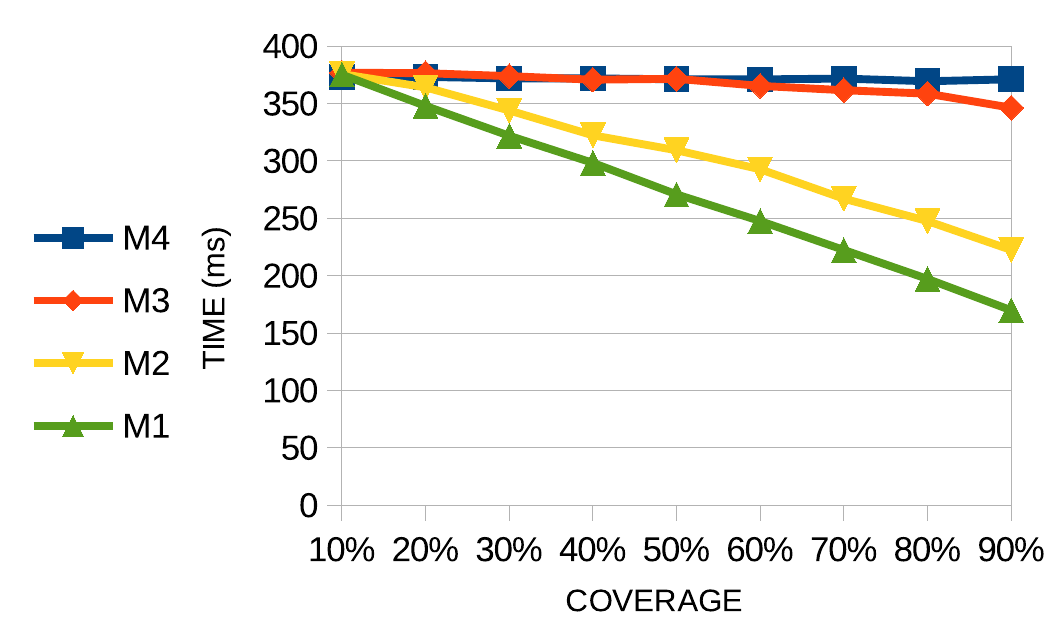}
        \caption{Small Size Model Query }
        \label{figure:small_query_vs_summary}
    \end{subfigure}%
    \begin{subfigure}[b]{0.3\linewidth}
        \centering
        \includegraphics[width=2.in,height=1.5in]{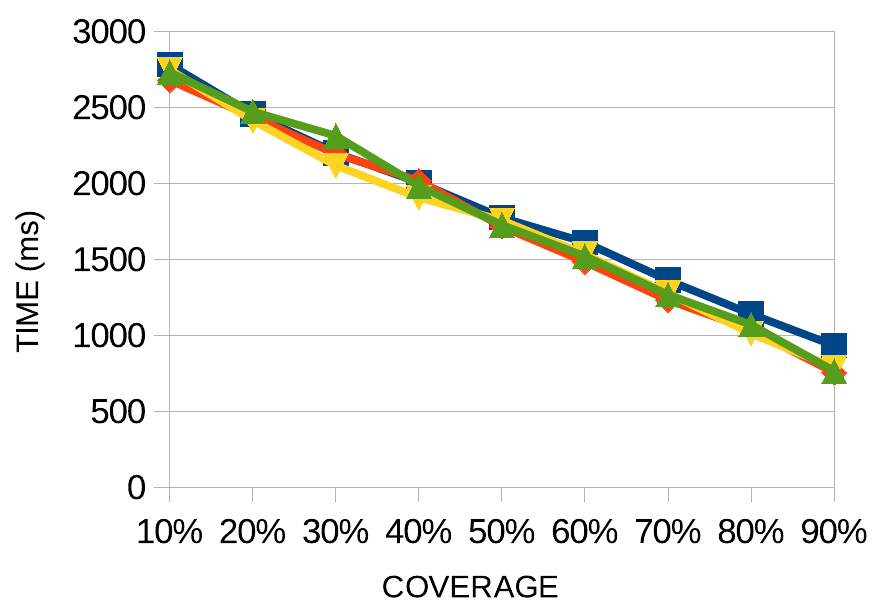}
        \caption{Large Size Model Query }
        \label{figure:large_query_vs_summary}
    \end{subfigure}%
    \begin{subfigure}[b]{0.3\linewidth}
        \centering
        \includegraphics[width=2in,height=1.5in]{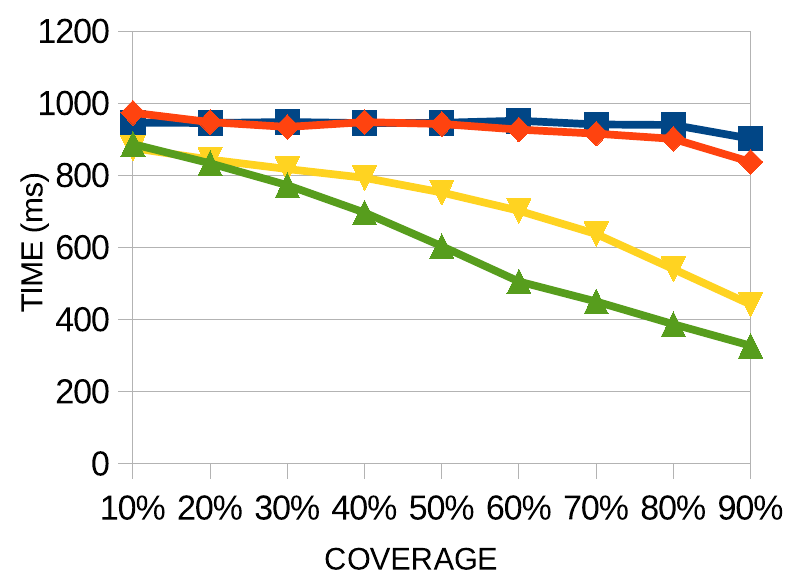}
        \caption{Small Size Model Query on Real Data}
        \label{figure:small_query_vs_summary_real}
    \end{subfigure}%
    \caption{Time take by large and small queries for various materialized model sizes}
    \label{figure:query_vs_summary}
 
 \end{figure*}
\subsection{Materialized Model Size and Performance Gain}

The size of materialized models is an important parameter in our framework. With the next set of experiments we wish to understand the impact of the size of materialized models on performance. Two test query sets S1 and S2 of size 50K  and 100k points are used as shown in the figure \ref{figure:summary_vs_speedup}. On the x-axis we represent different materialized model sets of fixed size of coverage fixed to 50\%. The size of the materialized model sets is varied from 5K points to 70K points as shown in the Figure \ref{figure:NB_summary_vs_speedup} and \ref{figure:LogR_summary_vs_speedup}. We present results for Naive Bayes (supports both incremental and decremental updates) and Logistic Regression (supports only incremental updates)  as similar trends hold for linear regression as well.  Figure \ref{figure:NB_summary_vs_speedup}, \ref{figure:LogR_summary_vs_speedup} present results for Naives Bayes and Logistic Regression respectively. We observe that for a fixed query size $D_q$ and fixed coverage there is an optimum size of materialized models which results in  maximum performance gain. We achieve a maximum performance gain for S1 at materialized model size of 20K for Naive Bayes. Similarly, for Logistic Regression we achieve the maximum performance gain at 10K materialized model size. As the size of the query increases the optimal materialized model size also increases. As shown in the graphs the query set S2 has its maximum at 30K and 20K for Naive Bayes and Logistic Regression respectively, which is larger than the maximum for S1. The exact position of the maximum on the graph depends on the size of the specific query (or query workload for multiple queries) for a given cost model.

\subsection{Materialized Model and Query Size}

We conducted experiments to quantify performance while scaling to larger input queries and materialized models sizes. The model chosen for these experiment was Naive Bayes, although linear regression also shows the same trends. Figure \ref{figure:query_vs_summary} shows four sizes of materialized models under consideration $M1$ to $M4$. $M1$ represents materialized models with their size chosen from a uniform distribution represented by U(25k,50k). Thus M1 is the scenario in which all the materialized models have a size uniformly 
distributed between 25K to 50K. Similarly M2,M3 and M4 are represented following a uniform distribution U(75k,100k), U(150K,200k) and U(250K,500K). Figure \ref{figure:small_query_vs_summary} shows the time taken to execute queries of small sizes represented by U(50K,100K). As depicted in the graph for M1 and M2 the time taken to execute the model queries decreases linearly as coverage increases. However for M3 and M4 which correspond to considerably larger materialized model sizes, the performance improvement becomes significant after 70\% coverage.  As coverage increases there is a higher probability to find two materialized models which can be subtracted in order to create a smaller model. Figure \ref{figure:small_query_vs_summary_real} shows similar trend for small queries on a real world data set from the UCI machine learning repository representing physical activity data of 3M points, consisting of 31 attributes and 13 classes. It is evident that the main trends
are the same as in the case of synthetic data set as is the case in all of our experiments. Figure \ref{figure:large_query_vs_summary} is the graph for larger query sizes represented by distribution U(500K,750K). Since the query size is much larger we can observe that all four cases materialized models are utilized to generate the model for the input query. For M1, small models can be combined to generate the models for larger data sets. While for M4 a large materialized model which has the maximum overlap with the incoming model construction query is manipulated to generate the new model. It is evident that the relationship of the query size to the materialized model size is
important in our setting. When the query workload has a much smaller size than the materialized model sizes (correspondingly when the query
workload has much larger size than the materialized model sizes) employing our framework does not result in large performance benefits. It
is evident however that enabling our framework in these cases does not impose an overhead either.

\begin{figure}    
  \centering
  \includegraphics[width=3in,height=1.5in]{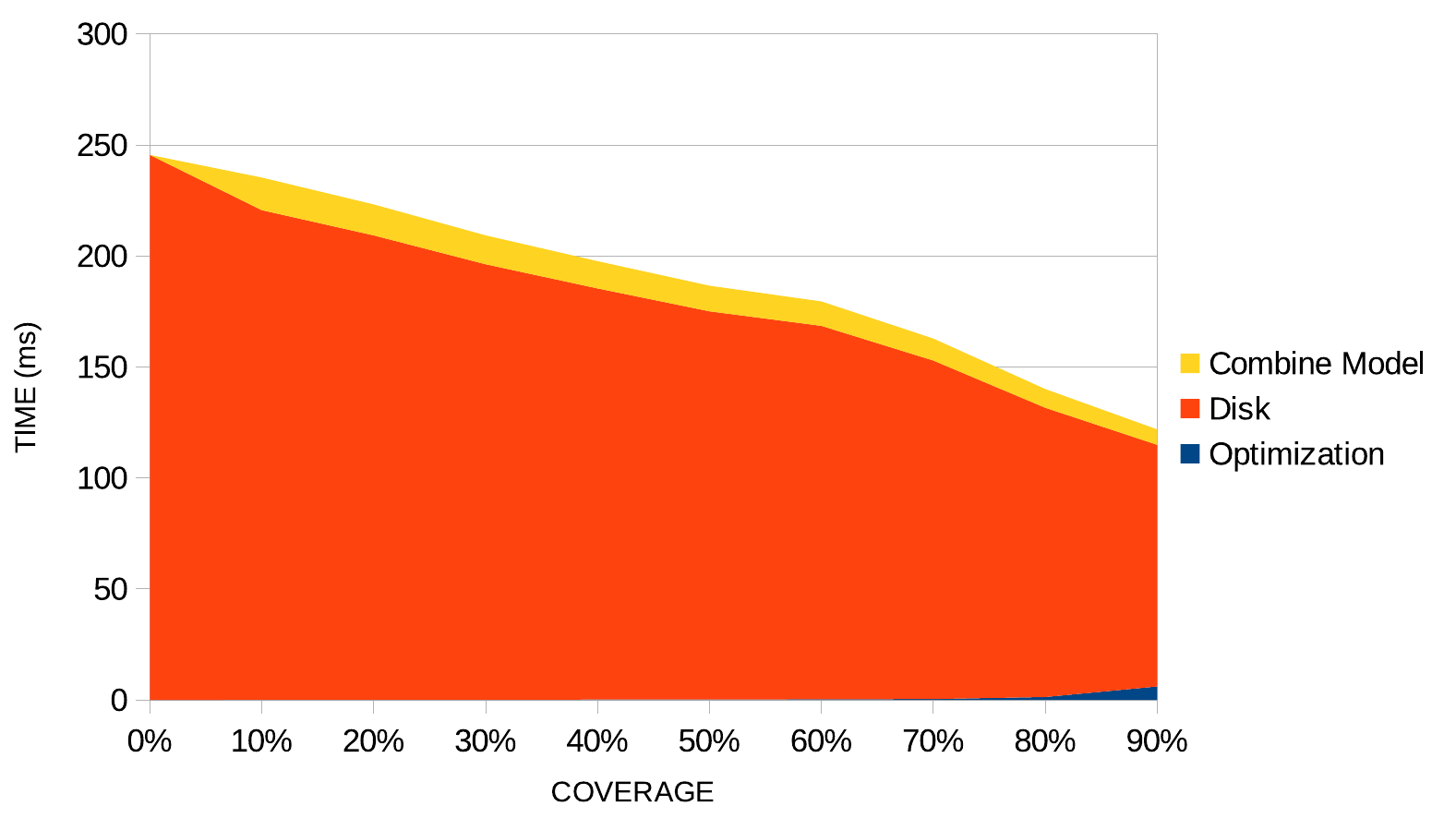}
  \caption{Distribution of time across various I/O and computation tasks}
  \label{figure:opt_vs_disk_time}
\end{figure}

\subsection{Optimization and I/O Time}

As mentioned in section \ref{sec:opt} the cost of merging models is considerably smaller as compared to disk access time. We measure the time taken by the three major components of our framework namely optimizer time, 
disk access time (including both fetching materialized model and/or fetching direct data points) and model combination time. The optimizer time refers to the time taken to run algorithm \ref{algo:optimization}. 
The time spend in fetching any information from MySQL is referred to as I/O time. The time remaining in our computations which cannot be attributed to the above cases is the time taken to merge the models. Experiments were run on a test set of a thousand queries. The size of the 
model to be generated is chosen from the normal distribution $\mathcal{N}(50K,12.5K)$.

\begin{figure*}
    
    \begin{subfigure}[b]{0.2\linewidth} 
        \centering
        \includegraphics[width=1.5in,height=1.5in]{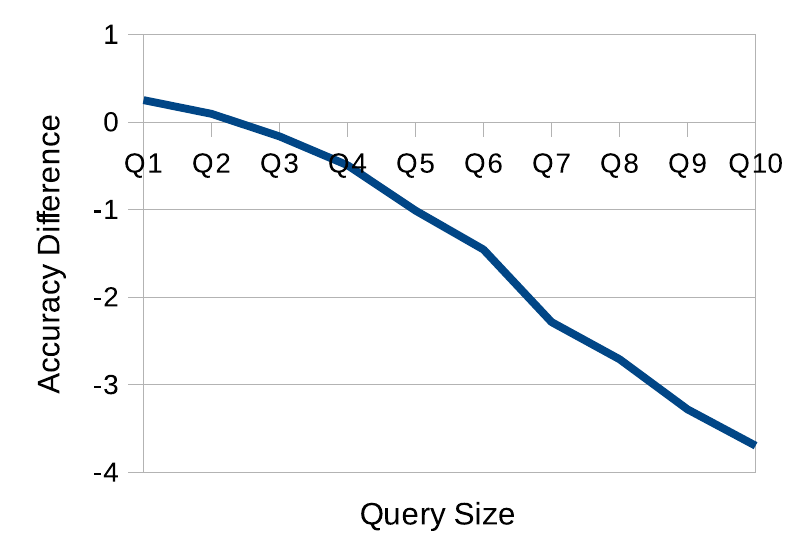}
        \caption{Average of diff. in accuracy}
        \label{figure:train_accuracy_avg_10k}
    \end{subfigure}%
   \hspace{1cm}
    \begin{subfigure}[b]{0.2\linewidth}
        \centering
        \includegraphics[width=1.5in,height=1.5in]{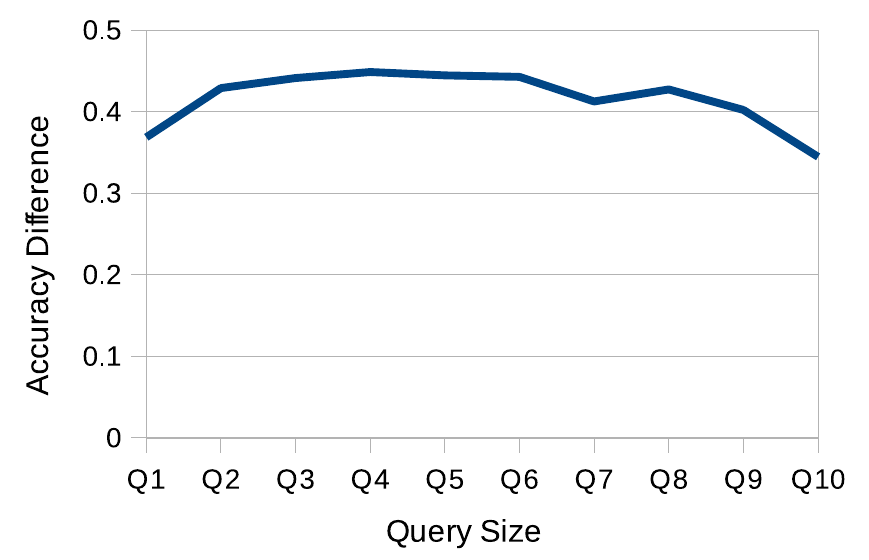}
        \caption{Average of positive diff. in accuracy}
        \label{figure:train_accuracy_pos_avg_10k}
    \end{subfigure}%
   \hspace{1cm}
    \begin{subfigure}[b]{0.2\linewidth}
        \centering
        \includegraphics[width=1.5in,height=1.5in]{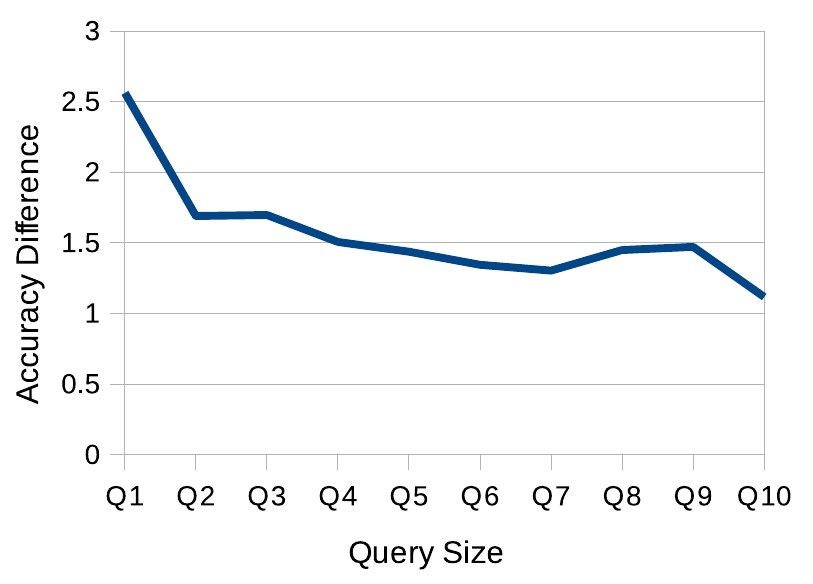}
        \caption{Maximum diff. in accuracy}
        \label{figure:train_accuracy_max_10k}
    \end{subfigure}
     \hspace{1cm}
     \begin{subfigure}[b]{0.2\linewidth}
        \centering
        \includegraphics[width=1.5in,height=1.5in]{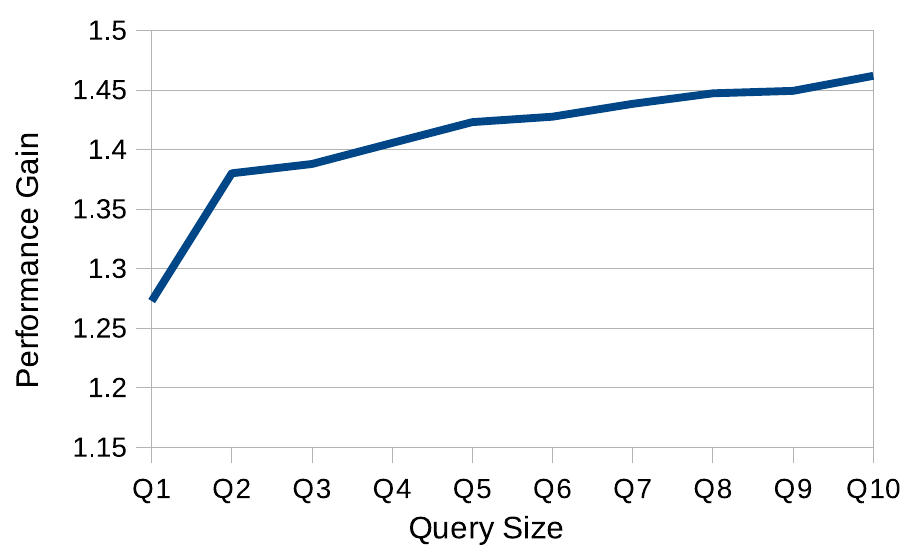}
        \caption{Performance Gain}
        \label{figure:train_accuracy_speedup_10k}
    \end{subfigure}
    \caption{Accuracy and Performance statistics for Logistic Regression with materialized model size of 10K}

    \begin{subfigure}[b]{0.2\linewidth}  
        \centering
        \includegraphics[width=1.5in,height=1.5in]{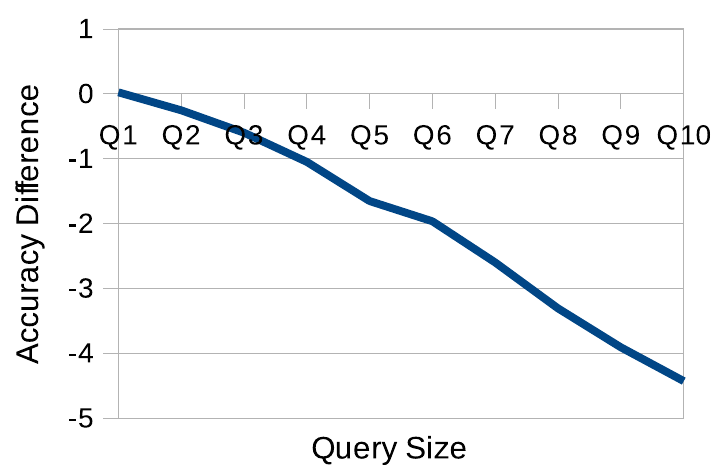}
        \caption{Average of diff. in accuracy}
        \label{figure:train_accuracy_avg_20k}
    \end{subfigure}%
   \hspace{1cm}
    \begin{subfigure}[b]{0.2\linewidth}
        \centering
        \includegraphics[width=1.5in,height=1.5in]{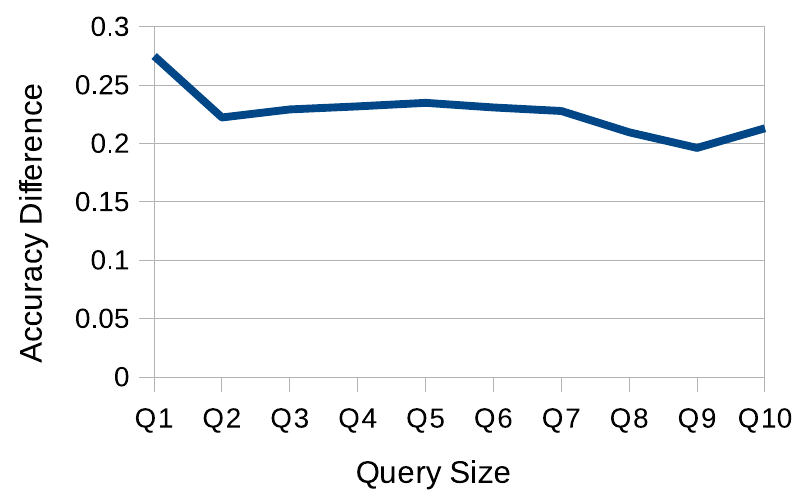}
        \caption{Average of positive diff. in accuracy}
        \label{figure:train_accuracy_pos_avg_20k}
    \end{subfigure}%
   \hspace{1cm}
    \begin{subfigure}[b]{0.2\linewidth}
        \centering
        \includegraphics[width=1.5in,height=1.5in]{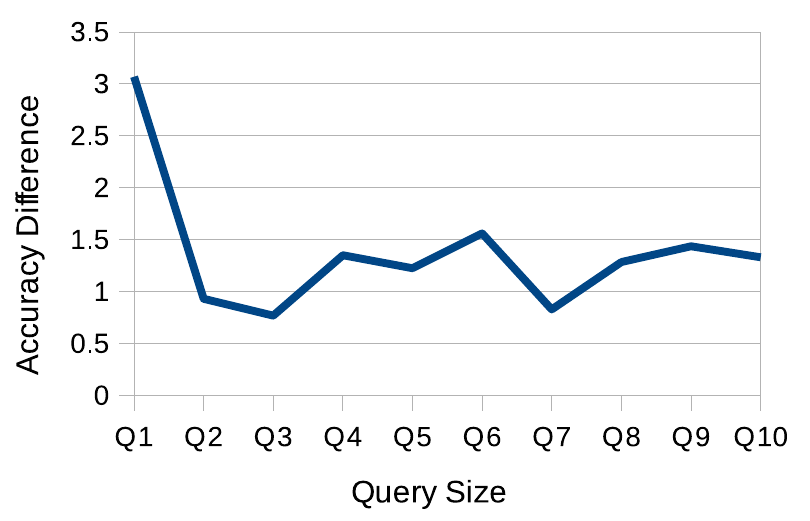}
        \caption{Maximum diff. in accuracy}
        \label{figure:train_accuracy_max_20k}
    \end{subfigure}
     \hspace{1cm}
     \begin{subfigure}[b]{0.2\linewidth}
        \centering
        \includegraphics[width=1.5in,height=1.5in]{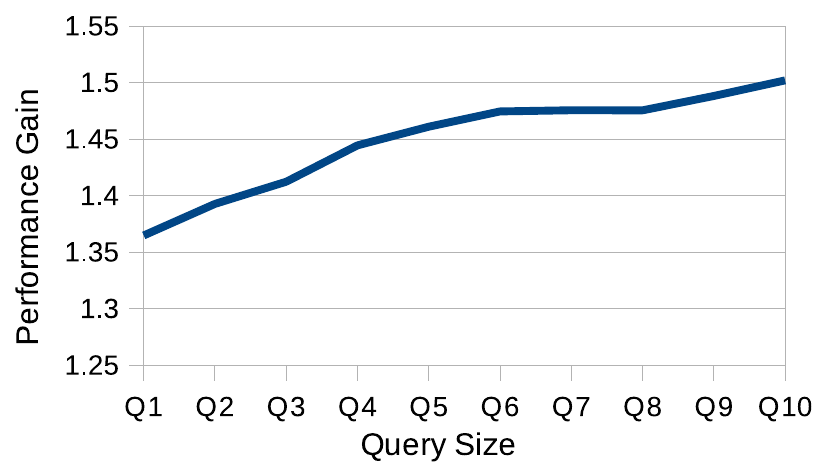}
        \caption{Performance Gain}
        \label{figure:train_accuracy_speedup_20k}
    \end{subfigure}
    \caption{Accuracy and Performance statistics for Logistic Regression with materialized model size of 20K}
    \label{figure:accuracy}
    
\end{figure*}

The expected time for each component is reported as shown in graph \ref{figure:opt_vs_disk_time}. As can be observed the majority of time to create models is spent while fetching data from disk. Model combination time is fairly constant and is much smaller as compared to disk time. Optimizer time is insignificant for small coverage and only becomes visible (but still negligible) on the graph when coverage is close to 80\% and above. As coverage increases the number of possible execution plans become considerably larger thus the optimizer takes much longer to build the graph and determine the shortest paths in the graph. This graph reveals that the overhead of running the optimization is minimal. Since the potential benefits of considering materialized models are significant, it is evident that if one chooses to materialize models, the performance overhead of the optimizer is negligible. Thus, running the optimizer, even if the decision is to employ the baseline, imposes minimal penalty in the query performance.
In the graph the baseline is represented  by the x-axis value at zero percent coverage. It can be seen that disk time reduces from 250 ms to 110 ms, while the optimizer time and model combination time are roughly 10ms. Thus, when the coverage is low, the overhead of the optimizer is so small that even when no materialized model can be utilized and the model has to be constructed from the baseline, the impact of the optimizer to the overall performance is immaterial as evident in Figure \ref{figure:opt_vs_disk_time}. At high coverage, the chances of utilizing materialized models are much higher. In that case, the small overhead of the optimizer is clearly compensated by the large savings in model construction time. 

\subsection{Accuracy}

In this section we analyze the accuracy of our framework for the logistic regression models presented in section \ref{sec:logreg}. We quantify the accuracy of the overall approach.

Synthetically generated classification data with 10 features and 2 classes were used to run test experiments. Similar trends hold when the number of classes increases, so we omit these experiments for brevity. We ran experiments on a test set S of a thousand queries. For each of these queries the model was built using our framework and also by applying SGD. We compare the accuracy on training data for both models by computing their difference. Let $A$ refer to the accuracy of the model built by our framework and $A_0$ refer to accuracy of SGD algorithm, the accuracy difference can be represented as $A_0 - A$.  Various statistics are reported on this difference. Figure \ref{figure:train_accuracy_avg_10k} and \ref{figure:train_accuracy_avg_20k} presents the average of the accuracy difference between the model constructed by our approach and the model constructed by SGD directly. The x-axis represents queries in increasing order of size. The graphs show negative average values which means that on average the model generated by our framework outperforms the model developed by SGD on training data. Also as the query size increases the expected performance of our model improves.
Figure \ref{figure:train_accuracy_pos_avg_10k} and \ref{figure:train_accuracy_pos_avg_20k} presents the average difference in accuracy for the cases where $(A_0 - A) > 0 $. It can be seen that the average positive difference  lies within 0.5\%. It is evident that the overall approach is highly accurate.
Across the materialized model sizes we observe that larger size has better accuracy as compared to smaller sizes. Finally Figures \ref{figure:train_accuracy_max_10k} and \ref{figure:train_accuracy_max_20k} present  the maximum difference  across various query sizes. The graph shows that as the query size increases the maximum difference  between the model computed by our framework and that computed by SGD decreases. It is visible from the graph that $max(A_0-A) < 3\%$ . 
The last set of graphs presents the trade off between accuracy and the corresponding performance gains achieved by our framework. As figures \ref{figure:train_accuracy_speedup_10k} and \ref{figure:train_accuracy_speedup_20k} suggest we experience a performance gain of 1.5x while we compromise accuracy by 3\% in the worst case. Similar results were observed on real world data sets including the PAMAP2 publicly available data set \cite{UCI:MLR}. Since they are consistent with what has been presented these results are omitted for brevity.

\section{Related Work} \label{sec:related}

There has been an ever increasing interest to integrate statistical and machine learning capabilities to data management systems. 
Several efforts have been made in academia and industry to address this demand. Major database vendors now support analytical capabilities 
on top their database engines : IBM's SystemML \cite{IBM:SystemML} , Oracle's ORE \cite{Oracle:ORE}, SAP HANA \cite{SAP:HANA}. However the integration is loose and does not support notions of model persistence or incremental computations.
In the open source community one can observe similar trends with MADLib \cite{Pivotal:MADLib} library support for Postgres. 
Other data platforms like Spark and Hadoop also support machine libraries as an external layer on top of their data processing system with MLLib \cite{Apache:MLLib} and Mahout \cite{Apache:Mahout} respectively. 
Such approaches either utilize an existing data management platform and deploy its extensions to provide analytics capabilities or
represent systems that can execute machine learning and statistical packages. See \cite{T1CondieMPW13} for a general overview of systems support for machine learning and statistical operations. Haloop \cite{Bu2010} and Dryad \cite{isard2007} are examples of systems that utilize a form of persistence in their operations to improve the execution of a graph data flow. Although related in spirit, the approach and goal of these systems is to improve the
performance of specific iterative graph data flow computations; they do not address the case of synthesizing a new model by extending and/or combining past models which is central in our approach.

Recent work \cite{Kumar:GLM} focused on pushing machine learning primitives inside a relational database engine. Our work is intended as 
 a middle layer between the data processing engine and the analytical computing language layer. We require awareness of
previous computations by collecting them and explore materialized models to build new models for the data. Our goal is to
explore natural work sharing opportunities that exist in a typical data analysis workload.

Materializing portion of computations with the intention of reuse has also been explored in the domain of feature selection \cite{Zhang:FeatureSelection} for machine learning tasks. Our work
however explores the incremental updates and reuse of model to build new models.

\section{Conclusions} \label{sec:conc}

\balance
In this paper we presented an approach that utilizes model materialization and incremental model reuse as
a first class citizen while processing data analytics workloads. Utilizing popular machine learning models
we demonstrated their incremental aspects and detailed an optimization methodology that determines the 
best way (in terms of performance) to build a given new model. We demonstrated that our apporach can achieve
significant savings in performance for new model construction while only imposing modest overheads in storage.

The work opens several avenues for future work. First there is a plethora of other models that are important
and can be considered in conjunction with our framework. Studying their incremental aspects and embedding
them into the same optimization framework is an interesting direction for future work. Incremental model
reuse for analytics is an important direction of research that blends nicely with the way current data management
systems build integrations to existing analytical packages. Our framework can be easily injected between the analytical
package and the RDBMS and recognize as well as handle all opportunities for improved performance. We are currently
building such as system based on the ideas presented herein in which we will report soon.

Finally, our focus in this paper has been in the case that a total ordering exists in the underlying data set. An interesting 
case is when such an ordering does not exist. In that case the model descriptors will be different as well as
the associated optimizations. Indeed our entire framework can be extended for this case as well and we will
be reporting on such extensions in our future work.

%
\bibliographystyle{abbrv}
\bibliography{applus}  
%
%
\balancecolumns


\end{document}